\documentclass[pra, showkeys,showpacs]{revtex4b5} 
\usepackage{amssymb}
\usepackage{amsmath}
\usepackage{graphicx} 
\usepackage{bm}
\usepackage{amsfonts}
\usepackage{psfrag}
\usepackage{hyperref}
\usepackage{natbib}

\begin{document}

\title{Effect of Thermal Fluctuations in the Resonance Fluorescence of
a Three-Level System}

\author{H. Couto} 
\email{couto@fisica.ufjf.br}
\affiliation{Universidade Federal de Juiz de Fora,\\ Instituto de
Ci\^{e}ncias Exatas, \\ Departamento de F\'{\i }sica, \\ 36033-330
Juiz de Fora, MG, Brazil}

\author{L. A. Amarante Ribeiro}
\email{lincoln.bh@terrra.com.br}
\affiliation{Universidade Federal de Minas Gerais,\\
Instituto de Ci\^{e}ncias Exatas, \\
Departamento de F\'{\i}sica\\31270-901 Belo Horizonte, MG, Brazil}
\date{\today}

\begin{abstract}
The effect of thermal fluctuations  in the resonance fluorescence of a
three-level system is studied. The damped three-level system is driven
by two  strong incident classical fields  near resonances frequencies.
The simulation  of a thermal bath  is obtained with a  large system of
harmonic oscillators  that represent the  normal modes of  the thermal
radiation  field.   The  time   evolution  of  the  fluorescent  light
intensities are obtained solving  by a iterative method the Heisenberg
equations of  motion in the integral  form. The results  show that the
time  development  of  the  intensity  of the  fluorescence  light  is
strongly affected  by the interaction  of the system with  the thermal
bath.  
\end{abstract}



\maketitle

\section{Introduction}

  In the  last fifteen years,  quantum jumps in three-level  systems has
been extensively studied \cite{bib1}-\cite{bib13}. The three-level
system is a versatile model that has been used to study, among others,
the  correlations   in  emission  of   photons  \cite{bib4},  resonant
fluorescence \cite{bib6},  operation of a  two-mode laser \cite{bib7},
coherent   pump   dynamics   \cite{bib8},  squeezing   properties   of
electromagnetic  field \cite{bib10},  electron  shelving \cite{bib12},
quantum  measurements theory \cite{bib14}.  The interest  on processes
involving atoms  with few energy  levels recently increased  even more
with the possibility of to study experimentally non-linear processes
with ion trapping.  

In this paper we consider the fluorescence resonance of a three-level
system with a coherent interaction with two driving fields and a incoherent
interaction with a thermal reservoir. When the driving fields are
turned on, the system is driven to a new non-equilibrium steady
state. If the driven fields are turned off, the system returns to the  
original equilibrium state with the thermal bath. We assume that the
atom is in a cavity where there is a quantized radiation field in
thermal equilibrium with the atom and the cavity walls at a certain
temperature. 
In our  model of a three-level  atom, the allowed  transitions are only
between the levels 1 and 2 and 1 and 3. The three-level system of this
kind is known as the $V$ configuration. We study the system driven by the
interaction  with two  electromagnetic fields  of  frequencies $\omega
_{1l}$  e $\omega  _{2l}$,  near, respectively,  the  $\omega _{1}$  e
$\omega  _{2}$   frequencies  characteristics  of   the  system.   The
three-level system  also interacts with a radiation  field, with which
one maintains in thermal equilibrium. We suppose that the normal modes
of  the radiation  field constitutes  the  thermal bath  at a  certain
temperature. We obtain  the  equations of  motion  of the  dynamical
operators in the Heisenberg formalism in the general form
   \begin{equation}
   \frac{dx}{dt}=-\beta x + A(t)
   \end{equation}

\noindent that may be recognized as equations that describes a damped system
subject to fluctuating forces, accordingly Langevin's theory of
Brownian motion\cite{bib22},\cite{bib23}. In the present problem,
these fluctuating forces are represented by the non-linear terms in
the equation of motion due to the interaction of the system with the
thermal bath when $T\neq 0\,K$. The role of the fluctuating forces is to
bring the system to the thermal equilibrium. The non-linearity of the
equations, caused by the saturation of the atomic transitions,
enhances the atom-bath interaction.

\par Even though the works about fluorescence resonance take in
account the  interaction of the resonant system with a thermal bath,
they in general assume the temperature of the bath as $T=0\,K$. Senitzky
\cite{bib17}, however, in a elegant way regards the effect of
thermal fluctuations in the resonance fluorescence of a two-level
atom. But Senitzky treats a more general model for the thermal
reservoir than we present here, and his mathematical treatment is more
involved. 

In this paper we prefer to use the equations of motion of the dynamical
operators to calculate the intensity of the fluorescent
light. Although this approach is equivalent to the density operator,
it has the advantage to facilitate the physical interpretation,
because it resembles with the classical treatment given to the
Brownian movement, in terms of the Langevin's equations.

This paper is organized as follows. In Sec. 2 we describe formally
the system and obtain the Heisenberg equations of motion in integral
form. In Sec. 3 we determine the solution of the system of equations
and apply it to the spontaneous emission to obtain the mean intensity
of the scattered fields in the fluorescence of the two excited
levels. In Sec. 4 we  discuss and interpret graphically the results of
the Sec. 3. Some details of the calculations are given in Appendix. 

\section{The Hamiltonian of the system}

We consider here the problem of an atom fixed in space, with three
levels and one electron, assuming that transitions occur only between
each excited level and the fundamental one. Thus, the three-level
system will be described with the help of the following operators
   \begin{eqnarray}
   R_{00} &=& |0\rangle\langle 0|, \label{op1}\\
   R_{11} &=& |1\rangle\langle 1|, \\
   R_{22} &=& |2\rangle\langle 2|, \\
   R_{01} &=& |0\rangle\langle 1|, \\
   R_{02} &=& |0\rangle\langle 2|, \label{op6}
   \end{eqnarray}
which obey the relation
   \[
   1=|0\rangle \langle 0|+|1\rangle \langle 1|+|2\rangle \langle 2|.
   \]

The fundamental level is represented by $|0\rangle $, and the excited
levels are represented by $|1\rangle $ and $|2\rangle $,
respectively. The commutation relations of the $R_{ij}$ operators are

   \begin{eqnarray}
   \lbrack R_{01},R_{01}^{\dagger }] &=&R_{00}-R_{11}, \\ \lbrack
   R_{02},R_{02}^{\dagger }] &=&R_{00}-R_{22}, \\ \lbrack
   R_{02},R_{02}^{\dagger }] &=&-R_{12}=0, \\ \lbrack
   R_{01},R_{02}^{\dagger }] &=&-R_{21}=0, \label{Comut}
   \end{eqnarray}
where we make the transition rates between the two excited levels
vanish.  The atom is illuminated with two polarized laser beams; each
beam has a frequency close to the characteristic frequency of each
excited level of the atom. We also assume that the light beams are
intense, and they will be treated classically. Within this point of
view, the atom may be assumed as a couple of electric dipoles
interacting with the electromagnets fields of the light beams. To
account the radiative damping, a thermal bath is simulated with a big
system of harmonic oscillators, that perform the role of the normal
modes of the thermal radiation field. The Hamiltonian is

\begin{equation}
H=H_{0}+H_{I},
\end{equation}
where

\begin{equation}
H_{0}=\hbar \omega _{1}R_{11}+\hbar \omega _{2}R_{22}+\hbar
\sum_{k}\omega _{k}^{\prime }b_{k}^{\dagger }b_{k}.
\end{equation}
$b_{k}$ and $b_{k}^{\dagger }$ are the operators corresponding to the
modes of the bath (annihilation and creation, respectively), that
satisfy

\begin{equation}
\lbrack b_{k},b_{k^{\prime }}^{\dagger }]=\delta _{kk^{\prime }}.
\end{equation}

The interaction Hamiltonian, in the rotating wave approximation, is
given by
\begin{equation}
H_{I}=\left\{ -\hbar R_{01}^{\dagger }\left( \lambda _{1}E_{1}^{\prime
}(t)+i\sum_{k}g_{1}(k)b_{k}\right) -\hbar R_{02}^{\dagger }\left(
\lambda _{2}E_{2}^{\prime }(t)+i\sum_{k}g_{2}\left( k\right)
b_{k}\right) \right\} +h.c.
\end{equation}
where $E_{1}^{\prime }(t)$ and $E_{2}^{\prime }(t)$ are the driven
fields,
with frequencies respectively close to $\omega _{1}$ and $\omega _{2}$. $\lambda_i$
and $g_i$ are coupling constants. The frequencies $%
\omega _{1}$ and $\omega _{2}$ are assumed quite different, so the
driven field tuned with one frequency will not excite electrons to the
level corresponding to the other one. The Heisenberg equations of
motion of the operators are
\begin{eqnarray}
\frac{dR_{00}}{dt} &=&\left\{ -iR_{10}\left( \lambda _{1}E_{1}^{\prime
}(t)+i\sum_{k}g_{1}(k)b_{k}\right) -\right.              \nonumber \\
&&\left. -iR_{20}\left( \lambda _{2}E_{2}^{\prime
}(t)+i\sum_{k}g_{2}(k)b_{k}\right) \right\} +h.c.      \label{R00} \\
\frac{dR_{11}}{dt} &=&\left\{ iR_{10}\left( \lambda _{1}E_{1}^{\prime
}(t)+i\sum_{k}g_{1}(k)b_{k}\right) \right\} +h.c.      \label{R11} \\
\frac{dR_{22}}{dt} &=&\left\{ iR_{20}\left( \lambda _{2}E_{2}^{\prime
}(t)+i\sum_{k}g_{2}(k)b_{k}\right) \right\} +h.c.      \label{R22} \\
\frac{dR_{01}}{dt} &=&-i\omega _{1}R_{01}+i(R_{00}-R_{11})\left(
\lambda _{1}E_{1}^{\prime
}(t)+i\sum_{k}g_{1}(k)b_{k}\right)                     \label{R01} \\ 
\frac{dR_{02}}{dt} &=&-i\omega
_{2}R_{02}+i(R_{00}-R_{22})\left( \lambda _{2}E_{2}^{\prime}(t)+i\sum_{k}g_{2}(k)b_{k}\right)\label{R02} \\ 
\frac{db_{k}}{dt}&=&-i\omega _{k}^{\prime 
}b_{k}+g_{1}^{*}(k)R_{01}+g_{2}^{*}(k)R_{02},          \label{bk} 
\end{eqnarray}
where $h.c$. means the hermitian conjugate. 

Let us define the $B_{1}(t)$ operator as
   \begin{equation}
   B_{1}(t)=\sum_{k}g_{1}\left( k\right) b_{k}(t),
   \end{equation}
and $B_{2}(t)$ as
\begin{equation}
B_{2}(t)=\sum_{k}g_{2}\left( k\right) b_{k}(t).
\end{equation}
Integrating the equation (\ref{bk}), we may write
\begin{eqnarray}
B_{1}(t)
&=&B_{01}(t)+\sum_{k}|g_{1}(k)|^{2}\int_{0}^{t}R_{01}(t^{\prime
})e^{i\omega _{k}^{\prime }\left( t^{\prime }-t\right) }dt^{\prime }+
\nonumber \\ &&+\sum_{k}g_{1}(k)g_{2}^{*}\left( k\right)
\int_{0}^{t}R_{02}(t^{\prime })e^{i\omega _{k}^{\prime }\left(
t^{\prime }-t\right) }dt^{\prime },
\label{opB1}
\end{eqnarray}
\begin{eqnarray}
B_{2}(t) &=&B_{02}(t)+\sum_{k}g_{2}(k)g_{1}^{*}\left( k\right)
\int_{0}^{t}R_{01}(t^{\prime })e^{i\omega _{k}^{\prime }\left(
t^{\prime }-t\right) }dt^{\prime }+ \nonumber \\
&&+\sum_{k}|g_{1}(k)|^{2}\int_{0}^{t}R_{02}(t^{\prime })e^{i\omega
_{k}^{\prime }\left( t^{\prime }-t\right) }dt^{\prime }, \label{opB2}
\end{eqnarray}
where
\begin{equation}
B_{01}\left( t\right) =\sum_{k}g_{1}\left( k\right) b_{k}\left(
0\right) e^{-i\omega _{k}^{\prime }t}, \label{B01}
\end{equation}
and
\begin{equation}
B_{02}\left( t\right) =\sum_{k}g_{2}\left( k\right) b_{k}\left(
0\right) e^{-i\omega _{k}^{\prime }t}.  \label{B02}
\end{equation}
We can now use the expressions (\ref{opB1}) and (\ref{opB2}) in the
equations (\ref {R00}) - (\ref{R02}), to eliminate the $b_{k}(t)$
variables of the thermal bath:
\begin{equation}
\frac{dR_{00}}{dt}=-iR_{10}\left( \lambda _{1}E_{1}^{\prime
}(t)+iB_{1}(t)\right) -iR_{20}\left( \lambda _{2}E_{2}^{\prime
}(t)+iB_{2}(t)\right) +h.c.
\end{equation}
\begin{eqnarray}
\frac{dR_{11}}{dt} &=&\left\{ iR_{10}\left( \lambda _{1}E_{1}^{\prime
}(t)+iB_{1}(t)\right) \right\} +h.c. \\ \frac{dR_{22}}{dt} &=&\left\{
iR_{20}\left( \lambda _{2}E_{2}^{\prime }(t)+iB_{2}(t)\right) \right\}
+h.c.
\end{eqnarray}
\begin{equation}
\frac{dR_{01}}{dt}=-i\omega _{1}R_{01}+i\left( R_{00}-R_{11}\right)
\left( \lambda _{1}E_{1}^{\prime }(t)+iB_{1}(t)\right) ,
\end{equation}
\begin{equation}
\frac{dR_{02}}{dt}=-i\omega _{2}R_{02}+i\left( R_{00}-R_{22}\right)
\left( \lambda _{2}E_{2}^{\prime }(t)+iB_{2}(t)\right).
\end{equation}
This is a set of non-linear differential equations that cannot be
exactly solved. The non-linearity of these equations is due to the
interaction between the three-level system and the radiation
fields. They are the Langevin equations for the system and  $B_1(t)$ and
$B_2(t)$, by analogy with the Brownian movement, are the random fluctuating forces acting on the atom. To solve it we
can try some approximation. As a first
approximation, let us suppose that the interaction is sufficiently
weak to be disregarded. Thus, the operators will evolve in time as
\begin{eqnarray}
R_{01}\left( t^{\prime }\right) &=&R_{01}\left( t\right) e^{-i\omega
_{1}\left( t^{\prime }-t\right) }, \\ R_{02}\left( t^{\prime }\right)
&=&R_{02}\left( t\right) e^{-i\omega _{2}\left( t^{\prime }-t\right)
}.
\end{eqnarray}
With  these adiabatic approximations,  the equations  (\ref{opB1}) and
(\ref{opB2}) becomes 
\begin{eqnarray}
B_{1}(t) &=&B_{01}(t)+\sum_{k}|g_{1}(k)|^{2}R_{01}(t)\int_{0}^{t}e^{i\left(
\omega _{k}^{\prime }-\omega _{1}\right) \left( t^{\prime }-t\right)
}dt^{\prime }+  \nonumber \\
&&+\sum_{k}g_{1}(k)g_{2}^{*}\left( k\right) R_{02}(t)\int_{0}^{t}e^{i(\omega
_{k}^{\prime }-\omega _{2})\left( t^{\prime }-t\right) }dt^{\prime },
\label{opB11}
\end{eqnarray}
\begin{eqnarray}
B_{2}(t) &=&B_{02}(t)+\sum_{k}g_{2}(k)g_{1}^{*}\left( k\right)
R_{01}(t)\int_{0}^{t}e^{i(\omega _{k}^{\prime }-\omega _{1})\left( t^{\prime
}-t\right) }dt^{\prime }+  \nonumber \\
&&+\sum_{k}|g_{2}(k)|^{2}R_{02}(t)\int_{0}^{t}e^{i\left( \omega _{k}^{\prime
}-\omega _{2}\right) \left( t^{\prime }-t\right) }dt^{\prime }.
\label{opB21}
\end{eqnarray}
Assuming that the reservoir modes are very close, the sums in eqs. (\ref
{opB11}) and (\ref{opB21}) may be substituted by integrations, where the
number of modes in the frequency interval $d\omega ^{\prime }$ is given by $%
\varrho (\omega ^{\prime })d\omega ^{\prime }$. With this consideration, and
using the known result \cite{bib15} 
\begin{equation}
\int_{0}^{\infty }e^{i(\omega ^{\prime }-\omega )(t^{\prime }-t)}dt^{\prime
}=-i\frac{\mathcal{P}}{\omega ^{\prime }-\omega }+\pi \delta (\omega
^{\prime }-\omega ),
\end{equation}
where $\mathcal{P}$ is the principal value of the integral, we may
write the expressions $B_{1}(t)$ and $B_{2}(t)$ as  
\begin{eqnarray}
B_{1}(t) &=&B_{01}(t) +\left( \pi \varrho \left( \omega
_{1}\right) g_{1}^{2}\left( \omega _{1}\right) -i\mathcal{P}\int d\omega
^{\prime }\frac{\varrho \left( \omega ^{\prime }\right) g_{1}^{2}\left(
\omega ^{\prime }\right) }{\omega ^{\prime }-\omega _{1}}\right) R_{01}(t)+ 
\nonumber \\
&&+\left(\pi \varrho (\omega _{2})g_{1}\left( \omega _{2}\right) g_{2}^{*}(\omega_{2})-i\mathcal{P}\int d\omega ^{\prime }\frac{\varrho \left( \omega ^{\prime }\right)g_{1}\left( \omega ^{\prime }\right) g_{2}^{*}(\omega ^{\prime })}{\omega^{\prime }-\omega _{2}}\right) R_{02}(t) , \label{opB12}
\end{eqnarray}
\begin{eqnarray}
B_{2}(t) &=&B_{02}(t) +\left( \pi \varrho \left( \omega _{2}\right) g_{2}^{2}\left( \omega_{2}\right) -i\mathcal{P}\int d\omega ^{\prime }\frac{\varrho \left( \omega^{\prime }\right) g_{2}^{2}\left( \omega ^{\prime }\right) }{\omega ^{\prime}-\omega _{2}}\right) R_{02}(t)+\nonumber \\
&&+\left(\pi \varrho (\omega _{1})g_{2}\left( \omega_{1}\right) g_{1}^{*}(\omega _{1})-i\mathcal{P}\int  d\omega  ^{\prime  }\frac{\varrho  \left(\omega^{\prime}\right)g_{2}\left(\omega^{\prime}\right) g_{1}^{*}(\omega^{\prime})}{\omega^{\prime}-\omega_{1}}\right)R_{01}(t).  \label{opB22}
\end{eqnarray}
Inserting now the expressions (\ref{opB12}) and (\ref{opB22}) in the
equations of motion (\ref{R11})-(\ref{R02}) we obtain
\begin{eqnarray}
\frac{dR_{11}}{dt} &=&i\lambda _{1}R_{10}(t)E_{1}^{\prime }\left( t\right)
-i\lambda _{1}^{*}R_{01}\left( t\right) E_{1}^{\prime *}\left( t\right)
-R_{10}\left( t\right) B_{01}\left( t\right) -  \nonumber \\
&&-B_{01}^{\dagger }\left( t\right) R_{01}\left( t\right) -k_{1}R_{11}\left(
t\right) ,  \label{dR11}
\end{eqnarray}
\begin{eqnarray}
\frac{dR_{22}}{dt} &=&i\lambda _{2}R_{20}(t)E_{2}^{\prime }\left( t\right)
-i\lambda _{2}^{*}R_{02}\left( t\right) E_{2}^{\prime *}\left( t\right)
-R_{20}\left( t\right) B_{02}\left( t\right) -  \nonumber \\
&&-B_{02}^{\dagger }\left( t\right) R_{02}\left( t\right) -k_{2}R_{22}\left(
t\right) ,  \label{dR22}
\end{eqnarray}
\begin{eqnarray}
\frac{dR_{01}}{dt}  &=&-i(\omega  _{01}-ik_{1}/2)R_{01}\left( t\right)
-\left(k_{12}\left( \omega _{2}\right) -i\delta \omega _{12}\right) R_{02}\left(t\right) +  \nonumber \\
&&+\left( R_{11}\left( t\right) -R_{00}\left( t\right) \right) \left(
B_{01}\left( t\right) -i\lambda _{1}E_{1}^{\prime }\left( t\right) \right) ,
\label{dR01}
\end{eqnarray}
\begin{eqnarray}
\frac{dR_{02}}{dt} &=&-i(\omega _{02}-ik_{2}/2)R_{02}\left( t\right) -\left(
k_{21}\left( \omega _{1}\right) -i\delta \omega _{21}\right) R_{01}\left(
t\right) +  \nonumber \\
&&+\left( R_{22}\left( t\right) -R_{00}\left( t\right) \right) \left(
B_{02}\left( t\right) -i\lambda _{2}E_{2}^{\prime }\left( t\right) \right).
\label{dR02}
\end{eqnarray}
In the above expressions we have used 
\begin{equation}
\omega _{01}=\omega _{1}-\mathcal{P}\int d\omega ^{\prime }\frac{\rho \left(
\omega ^{\prime }\right) g_{1}^{2}\left( \omega ^{\prime }\right) }{\omega
^{\prime }-\omega _{1}},
\end{equation}
\begin{equation}
\omega _{02}=\omega _{2}-\mathcal{P}\int d\omega ^{\prime }\frac{\rho \left(
\omega ^{\prime }\right) g_{2}^{2}\left( \omega ^{\prime }\right) }{\omega
^{\prime }-\omega _{2}},
\end{equation}
\begin{equation}
\frac{k_{1}}{2}=\pi \rho \left( \omega _{1}\right) g_{1}^{2}\left( \omega
_{1}\right) ,  \label{k1}
\end{equation}
\begin{equation}
\frac{k_{2}}{2}=\pi \rho \left( \omega _{2}\right) g_{2}^{2}\left( \omega
_{2}\right) ,  \label{k2}
\end{equation}
\begin{equation}
k_{12}\left( \omega _{2}\right) =\pi \rho \left( \omega _{2}\right)
g_{1}\left( \omega _{2}\right) g_{2}^{*}\left( \omega _{2}\right) ,
\end{equation}
\begin{equation}
k_{21}\left( \omega _{1}\right) =\pi \rho \left( \omega _{1}\right)
g_{2}\left( \omega _{1}\right) g_{1}^{*}\left( \omega _{1}\right) ,
\end{equation}
\begin{equation}
\delta \omega _{12}\left( \omega _{2}\right) =\mathcal{P}\int d\omega
^{\prime }\frac{\rho \left( \omega ^{\prime }\right) g_{1}\left( \omega
^{\prime }\right) g_{2}^{*}\left( \omega ^{\prime }\right) }{\omega ^{\prime
}-\omega _{2}},
\end{equation}
\begin{equation}
\delta \omega _{21}\left( \omega _{1}\right) =\mathcal{P}\int d\omega
^{\prime }\frac{\rho \left( \omega ^{\prime }\right) g_{2}\left( \omega
^{\prime }\right) g_{1}^{*}\left( \omega ^{\prime }\right) }{\omega ^{\prime
}-\omega _{1}}.
\end{equation}
The integral forms of the equations (\ref{dR11})-(\ref{dR02}) are 
\begin{eqnarray}
R_{11}\left( t\right)  &=&R_{11}\left( 0\right) e^{-k_{1}t}+i\lambda
_{1}\int_{0}^{t}R_{10}\left( t^{\prime }\right) E_{1}^{\prime }\left(
t^{\prime }\right) e^{k_{1}\left( t^{\prime }-t\right) }dt^{\prime }- 
\nonumber \\
&&-i\lambda _{1}^{*}\int_{0}^{t}R_{01}\left( t^{\prime }\right)
E_{1}^{\prime *}\left( t^{\prime }\right) e^{k_{1}\left( t^{\prime
}-t\right) }dt^{\prime }-  \nonumber \\
&&-\int_{0}^{t}R_{10}\left( t^{\prime }\right) B_{01}\left( t^{\prime
}\right) e^{k_{1}\left( t^{\prime }-t\right) }dt^{\prime }-  \nonumber \\
&&-\int_{0}^{t}B_{01}^{\dagger }\left( t^{\prime }\right) R_{01}\left(
t^{\prime }\right) e^{k_{1}\left( t^{\prime }-t\right) }dt^{\prime }
\label{iR11}
\end{eqnarray}
\begin{eqnarray}
R_{22}\left( t\right)  &=&R_{22}\left( 0\right) e^{-k_{2}t}+i\lambda
_{2}\int_{0}^{t}R_{20}\left( t^{\prime }\right) E_{2}^{\prime }\left(
t^{\prime }\right) e^{k_{2}\left( t^{\prime }-t\right) }dt^{\prime }- 
\nonumber \\
&&-i\lambda _{2}^{*}\int_{0}^{t}R_{02}\left( t^{\prime }\right)
E_{2}^{\prime *}\left( t^{\prime }\right) e^{k_{2}\left( t^{\prime
}-t\right) }dt^{\prime }-  \nonumber \\
&&-\int_{0}^{t}R_{20}\left( t^{\prime }\right) B_{02}\left( t^{\prime
}\right) e^{k_{2}\left( t^{\prime }-t\right) }dt^{\prime }-  \nonumber \\
&&-\int_{0}^{t}B_{02}^{\dagger }\left( t^{\prime }\right) R_{02}\left(
t^{\prime }\right) e^{k_{2}\left( t^{\prime }-t\right) }dt^{\prime }
\label{iR22}
\end{eqnarray}
\begin{eqnarray} 
R_{01}\left(t\right)&=&R_{01}\left(0\right)e^{-i\left(\omega_{01}-ik_{1}/2\right) t}- \nonumber \\
&&-\left( k_{12}\left( \omega _{2}\right) -i\delta \omega _{12}\right)
\int_{0}^{t}R_{02}\left( t^{\prime }\right) e^{i\left( \omega
_{01}-ik_{1}/2\right)  \left(   t^{\prime  }-t\right)  }dt^{\prime  }+
\nonumber \\
&&+\int_{0}^{t}\left( R_{11}\left( t^{\prime }\right) -R_{00}\left(
t^{\prime }\right) \right) \left( B_{01}\left( t^{\prime }\right) -i\lambda
_{1}E_{1}^{\prime }\left( t^{\prime }\right) \right) e^{i\left( \omega
_{01}-ik_{1}/2\right)   \left(  t^{\prime  }-t\right)   }dt^{\prime  }
\nonumber \\
&& \label{iR01}
\end{eqnarray}

\begin{eqnarray}
R_{02}\left( t\right)  &=&R_{02}\left( 0\right) e^{-i\left( \omega
_{02}-ik_{2}/2\right) t}-  \nonumber \\
&&-\left( k_{21}\left( \omega _{1}\right) -i\delta \omega _{21}\right)
\int_{0}^{t}R_{01}\left( t^{\prime }\right) e^{i\left( \omega
_{02}-ik_{2}/2\right) \left( t^{\prime }-t\right) }dt^{\prime }+  \nonumber
\\
&&+\int_{0}^{t}\left( R_{22}\left( t^{\prime }\right) -R_{00}\left(
t^{\prime }\right) \right) \left( B_{02}\left( t^{\prime }\right) -i\lambda
_{2}E_{2}^{\prime }\left( t^{\prime }\right) \right) e^{i\left( \omega
_{02}-ik_{2}/2\right) \left( t^{\prime }-t\right) }dt^{\prime }  \nonumber \\
&&  \label{iR02}
\end{eqnarray}
Supposing that the external fields are monochromatic and
plane-polarized, we write   
\begin{equation}
E_{1}^{\prime }\left( t\right) =E_{01}e^{-i\omega _{1l}t},
\end{equation}
\begin{equation}
E_{2}^{\prime }\left( t\right) =E_{02}e^{-i\omega _{2l}t}.
\end{equation}
Introducing this form of the external fields in eqs. (\ref{iR11})-(\ref{iR02}%
), and taking the mean values, we obtain 
\begin{eqnarray}
\langle R_{11}\left( t\right) \rangle  &=&\langle R_{11}\left( 0\right)
\rangle e^{-k_{1}t}+i\lambda _{1}E_{01}\int_{0}^{t}\langle R_{10}\left(
t^{\prime }\right) \rangle e^{-i\omega _{1l}t^{\prime }}e^{k_{1}\left(
t^{\prime }-t\right) }dt^{\prime }-  \nonumber \\
&&-i\lambda _{1}^{*}E_{01}^{*}\int_{0}^{t}\langle R_{01}\left( t^{\prime
}\right) \rangle e^{-i\omega _{1l}t^{\prime }}e^{k_{1}\left( t^{\prime
}-t\right) }dt^{\prime }-  \nonumber \\
&&-\int_{0}^{t}\langle R_{10}\left( t^{\prime }\right) B_{01}\left(
t^{\prime }\right) \rangle e^{k_{1}\left( t^{\prime }-t\right) }dt^{\prime }-
\nonumber \\
&&-\int_{0}^{t}\langle B_{01}^{\dagger }\left( t^{\prime }\right)
R_{01}\left( t^{\prime }\right) \rangle e^{k_{1}\left( t^{\prime }-t\right)
}dt^{\prime },  \label{R11M}
\end{eqnarray}
\begin{eqnarray}
\langle R_{22}\left( t\right) \rangle  &=&\langle R_{22}\left( 0\right)
\rangle e^{-k_{2}t}+i\lambda _{2}E_{02}\int_{0}^{t}\langle R_{20}\left(
t^{\prime }\right) \rangle e^{-i\omega _{2l}t^{\prime }}e^{k_{2}\left(
t^{\prime }-t\right) }dt^{\prime }-  \nonumber \\
&&-i\lambda _{2}^{*}E_{02}^{*}\int_{0}^{t}\langle R_{02}\left( t^{\prime
}\right) \rangle e^{-i\omega _{2l}t^{\prime }}e^{k_{2}\left( t^{\prime
}-t\right) }dt^{\prime }-  \nonumber \\
&&-\int_{0}^{t}\langle R_{20}\left( t^{\prime }\right) B_{02}\left(
t^{\prime }\right) \rangle e^{k_{2}\left( t^{\prime }-t\right) }dt^{\prime }-
\nonumber \\
&&-\int_{0}^{t}\langle B_{02}^{\dagger }\left( t^{\prime }\right)
R_{02}\left( t^{\prime }\right) \rangle e^{k_{2}\left( t^{\prime }-t\right)
}dt^{\prime },  \label{R22M}
\end{eqnarray}
\begin{eqnarray}
\langle R_{01}\left( t\right) \rangle  &=&\langle R_{01}\left( 0\right)
\rangle e^{-i\left( \omega _{01}-ik_{1}/2\right) t}-  \nonumber \\
&&-\left( k_{12}\left( \omega _{2}\right) -i\delta \omega _{12}\left( \omega
_{2}\right) \right) \int_{0}^{t}\langle R_{02}\left( t^{\prime }\right)
\rangle e^{i\left( \omega _{01}-ik_{1}/2\right) \left( t^{\prime }-t\right)
}dt^{\prime }+  \nonumber \\
&&+\int_{0}^{t}\left\langle \left( 2R_{11}(t^{\prime })+R_{22}\left(
t^{\prime }\right) -1\right) B_{01}\left( t^{\prime }\right) \right\rangle
e^{i\left( \omega _{01}-ik_{1}/2\right) \left( t^{\prime }-t\right)
}dt^{\prime }+  \nonumber \\
&&-i\lambda _{1}E_{01}\int_{0}^{t}\left\langle \left( 2R_{11}\left(
t^{\prime }\right) +R_{22}\left( t^{\prime }\right) -1\right) \right\rangle
e^{-i\omega _{1l}t^{\prime }}e^{i\left( \omega _{01}-ik_{1}/2\right) \left(
t^{\prime }-t\right) }dt^{\prime },  \nonumber \\
&&  \label{R01M}
\end{eqnarray}
\begin{eqnarray}
\langle R_{02}\left( t\right) \rangle  &=&\langle R_{02}\left( 0\right)
\rangle e^{-i\left( \omega _{02}-ik_{2}/2\right) t}-  \nonumber \\
&&-\left( k_{21}\left( \omega _{1}\right) -i\delta \omega _{21}\left( \omega
_{1}\right) \right) \int_{0}^{t}\langle R_{01}\left( t^{\prime }\right)
\rangle e^{i\left( \omega _{02}-ik_{2}/2\right) \left( t^{\prime }-t\right)
}dt^{\prime }+  \nonumber \\
&&+\int_{0}^{t}\langle \left( 2R_{22}(t^{\prime })+R_{11}\left( t^{\prime
}\right) -1\right) B_{02}\left( t^{\prime }\right) \rangle e^{i\left( \omega
_{02}-ik_{2}/2\right) \left( t^{\prime }-t\right) }dt^{\prime }+  \nonumber
\\
&&-i\lambda _{2}E_{02}\int_{0}^{t}\langle \left( 2R_{22}\left( t^{\prime
}\right) +R_{11}\left( t^{\prime }\right) -1\right) \rangle e^{-i\omega
_{2l}t^{\prime }}e^{i\left( \omega _{02}-ik_{2}/2\right) \left( t^{\prime
}-t\right) }dt^{\prime }.  \nonumber \\
&&  \label{R02M}
\end{eqnarray}

\section{Spontaneous emission and resonant fluorescence}

We follow the procedure outlined in ref. \cite{bib16} to obtain the solution
of the system of integral equations (\ref{R11M}-\ref{R02M}). We use a
iterative process: the expressions $R_{11}$, $R_{22}$, $R_{01}$ and $R_{02}$
are linear in the initial values of the operators and the coefficients of
the operators are expressed in terms of powers of $B_{01}(t)$ and $B_{02}(t)$%
. Supposing that the radiation field is in a stationary state and has a
Gaussian distribution, the higher-order correlation function can be
expressed in terms of the second-order one\cite{bib16}, 
\begin{equation}
\langle B_{0}^{\dagger }(t_{1})\cdot \cdot \cdot B_{0}^{\dagger
}(t_{n})B_{0}(t_{n}^{\prime })\cdot \cdot \cdot B_{0}(t_{1}^{\prime
})\rangle =\sum_{\left| j\right| }\prod_{i=1}^{n}\left\langle B_{0}^{\dagger
}(t_{j})B_{0}(t_{i}^{\prime })\right\rangle.  \label{corrl}
\end{equation}
The sum must be taken over all permutations $j=1,\ldots,n$. As the expressions
of the operators depend on the radiations field configuration in earlier
times, the only terms that contribute to the series expansion of equations (%
\ref{R11M}) and (\ref{R22M}) are those for which the field operators are
time ordered, 
\[
\text{min}\left( t_{i},t_{i}^{\prime }\right) \geq \text{ max}\left(
t_{i+1},t_{i+1}^{\prime }\right) \text{, \qquad }i=1,2,\cdot \cdot \cdot ,n. 
\]
The only term in the sum of the equation (\ref{corrl}) that contributes is 
\[
\prod_{i=1}^{n}\left\langle B_{0}^{\dagger }(t_{j})B_{0}(t_{i}^{\prime
})\right\rangle. 
\]
Thus, terms that contain a different number of factors $B_{0}^{\dagger }(t)$
and $B_{0}(t)$ do not contribute to the mean value of the operator. A
further simplification can be obtained if we suppose that the radiation
field spectrum is dense, flat and broad. In this case, the correlation
functions $\left\langle B_{10}^{\dagger }(t^{\prime \prime
})B_{10}(t_{i}^{\prime })\right\rangle $ and $\left\langle B_{20}^{\dagger
}(t^{\prime \prime })B_{20}(t_{i}^{\prime })\right\rangle $ can be
calculated, by using equations (\ref{B01}) and (\ref{B02}), 
\begin{equation}
\left\langle B_{10}^{\dagger }(t^{\prime \prime })B_{10}(t_{i}^{\prime
})\right\rangle =\sum_{k}\left| g_{1}\left( k\right) \right| ^{2}N\left(
\omega \right) e^{i\left( \omega _{k}-\omega _{01}\right) \left( t^{\prime
\prime }-t^{\prime }\right) },  \label{b1b1}
\end{equation}
\begin{equation}
\left\langle B_{20}^{\dagger }(t^{\prime \prime })B_{20}(t_{i}^{\prime
})\right\rangle =\sum_{k}\left| g_{2}\left( k\right) \right| ^{2}N\left(
\omega \right) e^{i\left( \omega _{k}-\omega _{02}\right) \left( t^{\prime
\prime }-t^{\prime }\right) },  \label{b2b2}
\end{equation}
where 
\begin{equation}
N\left( \omega \right) =\langle b_{k}^{\dagger }(0)b_{k}\left( 0\right)
\rangle
\end{equation}
is the initial mean number of photons at the $k$ mode of the radiation
field. Substituting the sums in equations (\ref{b1b1}) and (\ref{b2b2}) by
integrations, we have 
\begin{equation}
\left\langle B_{10}^{\dagger }(t^{\prime \prime })B_{10}(t_{i}^{\prime
})\right\rangle =\int_{0}^{\infty }d\omega \rho (\omega )g_{1}^{2}\left(
\omega \right) N\left( \omega \right) e^{i\left( \omega _{k}-\omega
_{01}\right) \left( t^{\prime \prime }-t^{\prime }\right) },
\end{equation}
\begin{equation}
\left\langle B_{20}^{\dagger }(t^{\prime \prime })B_{20}(t_{i}^{\prime
})\right\rangle =\int_{0}^{\infty }d\omega \rho (\omega )g_{2}^{2}\left(
\omega \right) N\left( \omega \right) e^{i\left( \omega _{k}-\omega
_{02}\right) \left( t^{\prime \prime }-t^{\prime }\right) },
\end{equation}
and using the properties of the Dirac delta-function and definitions (\ref
{k1}) and (\ref{k2}), we obtain 
\begin{equation}
\langle B_{10}^{\dagger }(t^{\prime \prime })B_{10}(t^{\prime })\rangle =%
\frac{k_{1}}{2}N(\omega _{1})\delta (t^{\prime \prime }-t^{\prime }),
\label{b1}
\end{equation}
\begin{equation}
\langle B_{20}^{\dagger }(t^{\prime \prime })B_{20}(t^{\prime })\rangle =%
\frac{k_{2}}{2}N(\omega _{2})\delta (t^{\prime \prime }-t^{\prime }).
\label{b2}
\end{equation}
Now we can calculate the correlations that appear in eqs. (\ref{R11M}) and (%
\ref{R22M}). Using relations (\ref{corrl}), (\ref{b1}) and (\ref{b2}), in
expressions (\ref{iR11}) and (\ref{iR22}), we obtain the expressions of $%
\langle R_{10}(t^{\prime })B_{01}(t^{\prime })\rangle $ and $\langle
R_{20}(t^{\prime })B_{02}(t^{\prime })\rangle $: 
\begin{equation}
\langle R_{10}(t^{\prime })B_{01}(t^{\prime })\rangle =k_{1}N(\omega
_{1})\langle R_{11}(t^{\prime })\rangle +\frac{k_{1}}{2}\left( \langle
R_{22}(t^{\prime })\rangle -1\right) ,  \label{R1B}
\end{equation}
\begin{equation}
\langle R_{20}(t^{\prime })B_{02}(t^{\prime })\rangle =k_{2}N(\omega
_{2})\langle R_{22}(t^{\prime })\rangle +\frac{k_{2}}{2}\left( \langle
R_{11}(t^{\prime })\rangle -1\right).  \label{R2B}
\end{equation}
Expressions (\ref{R1B}) and (\ref{R2B}) are real, and then, 
\begin{equation}
\langle R_{10}(t^{\prime })B_{01}(t^{\prime })\rangle =\langle
B_{01}^{\dagger }(t^{\prime })R_{01}(t^{\prime })\rangle ,
\end{equation}
\begin{equation}
\langle R_{20}(t^{\prime })B_{02}(t^{\prime })\rangle =\langle
B_{02}^{\dagger }(t^{\prime })R_{02}(t^{\prime })\rangle.
\end{equation}
Inserting eqs. (\ref{R1B}) and the complex conjugate (\ref{R01M}) in eq. (%
\ref{R11M}), we obtain 
\begin{eqnarray}
\langle R_{11}(t)\rangle &=&\langle R_{11}(0)\rangle e^{-k_{1}t}+\Gamma
_{21}(e^{-z_{1}t}-e^{-k_{1}t})+\Gamma _{21}^{*}(e^{-z_{1}^{*}t}-e^{-k_{1}t})+
\nonumber \\
&&\Gamma _{11}(1-e^{-k_{1}t})-2k_{1}N(\omega _{1})\int_{0}^{t}\langle
R_{11}(t^{\prime })\rangle e^{k_{1}(t^{\prime }-t)}dt^{\prime }-  \nonumber
\\
&&-k_{1}N(\omega _{1})\int_{0}^{t}\langle R_{22}(t^{\prime })\rangle
e^{k_{1}(t^{\prime }-t)}dt^{\prime }-  \nonumber \\
&&-\frac{\Omega _{1}^{2}}{2}\int_{0}^{t}dt^{\prime
}e^{-k_{1}t}e^{z_{1}^{*}t^{\prime }}\left\{ \int_{0}^{t^{\prime }}\langle
R_{11}(t^{\prime \prime })\rangle e^{z_{1}t^{^{\prime \prime }}}dt^{\prime
\prime }\right\} -  \nonumber \\
&&-\frac{\Omega _{1}^{2}}{2}\int_{0}^{t}dt^{\prime
}e^{-k_{1}t}e^{z_{1}t^{\prime }}\left\{ \int_{0}^{t^{\prime }}\langle
R_{11}(t^{\prime \prime })\rangle e^{z_{1}^{*}t^{^{\prime \prime
}}}dt^{\prime \prime }\right\} -  \nonumber \\
&&-\frac{\Omega _{1}^{2}}{4}\int_{0}^{t}dt^{\prime
}e^{-k_{1}t}e^{z_{1}^{*}t^{\prime }}\left\{ \int_{0}^{t^{\prime }}\langle
R_{22}(t^{\prime \prime })\rangle e^{z_{1}t^{^{\prime \prime }}}dt^{\prime
\prime }\right\} -  \nonumber \\
&&-\frac{\Omega _{1}^{2}}{4}\int_{0}^{t}dt^{\prime
}e^{-k_{1}t}e^{z_{1}t^{\prime }}\left\{ \int_{0}^{t^{\prime }}\langle
R_{22}(t^{\prime \prime })\rangle e^{z_{1}^{*}t^{^{\prime \prime
}}}dt^{\prime \prime }\right\}.  \label{R11M2}
\end{eqnarray}
Inserting the equation (\ref{R2B}) and the complex conjugate of eq. (\ref
{R02M}) in eq. (\ref{R22M}), we have 
\begin{eqnarray}
\langle R_{22}(t)\rangle &=&\langle R_{22}(0)\rangle e^{-k_{2}t}+\Gamma
_{22}(e^{-z_{2}t}-e^{-k_{2}t})+\Gamma _{22}^{*}(e^{-z_{2}^{*}t}-e^{-k_{2}t})+
\nonumber \\
&&\Gamma _{12}(1-e^{-k_{2}t})-2k_{2}N(\omega _{2})\int_{0}^{t}\langle
R_{22}(t^{\prime })\rangle e^{k_{2}(t^{\prime }-t)}dt^{\prime }-  \nonumber
\\
&&-k_{2}N(\omega _{2})\int_{0}^{t}\langle R_{11}(t^{\prime })\rangle
e^{k_{2}(t^{\prime }-t)}dt^{\prime }-  \nonumber \\
&&-\frac{\Omega _{2}^{2}}{2}\int_{0}^{t}dt^{\prime
}e^{-k_{2}t}e^{z_{2}^{*}t^{\prime }}\left\{ \int_{0}^{t^{\prime }}\langle
R_{22}(t^{\prime \prime })\rangle e^{z_{2}t^{^{\prime \prime }}}dt^{\prime
\prime }\right\} -  \nonumber \\
&&-\frac{\Omega _{2}^{2}}{2}\int_{0}^{t}dt^{\prime
}e^{-k_{2}t}e^{z_{2}t^{\prime }}\left\{ \int_{0}^{t^{\prime }}\langle
R_{22}(t^{\prime \prime })\rangle e^{z_{2}^{*}t^{^{\prime \prime
}}}dt^{\prime \prime }\right\} -  \nonumber \\
&&-\frac{\Omega _{2}^{2}}{4}\int_{0}^{t}dt^{\prime
}e^{-k_{2}t}e^{z_{2}^{*}t^{\prime }}\left\{ \int_{0}^{t^{\prime }}\langle
R_{11}(t^{\prime \prime })\rangle e^{z_{2}t^{^{\prime \prime }}}dt^{\prime
\prime }\right\} -  \nonumber \\
&&-\frac{\Omega _{2}^{2}}{4}\int_{0}^{t}dt^{\prime
}e^{-k_{2}t}e^{z_{2}t^{\prime }}\left\{ \int_{0}^{t^{\prime }}\langle
R_{11}(t^{\prime \prime })\rangle e^{z_{2}^{*}t^{^{\prime \prime
}}}dt^{\prime \prime }\right\}.  \label{R22M2}
\end{eqnarray}
In expressions (\ref{R11M2}) and (\ref{R22M2}) we use the below definitions 
\begin{equation}
\Gamma _{11}=N(\omega _{1})+\frac{\Omega _{1}^{2}}{4\left| z_{1}\right| ^{2}}%
,
\end{equation}
\begin{equation}
\Gamma _{12}=N(\omega _{2})+\frac{\Omega _{2}^{2}}{4\left| z_{2}\right| ^{2}}%
,
\end{equation}
\begin{equation}
\Gamma _{21}=\frac{i\lambda _{1}E_{01}\langle R_{10}(0)\rangle }{z_{1}^{*}}-%
\frac{\Omega _{1}^{2}}{4\left| z_{1}\right| ^{2}},
\end{equation}
\begin{equation}
\Gamma _{22}=\frac{i\lambda _{2}E_{02}\langle R_{20}(0)\rangle }{z_{2}^{*}}-%
\frac{\Omega _{2}^{2}}{4\left| z_{2}\right| ^{2}},
\end{equation}
\begin{eqnarray}
\Omega _{1} &=&2\left| \lambda _{1}\right| \left| E_{01}\right| , \\
\Omega _{2} &=&2\left| \lambda _{2}\right| \left| E_{02}\right| ,
\end{eqnarray}
\begin{equation}
z_{1}=\frac{k_{1}}{2}+i(\omega _{1l}-\omega _{01}),
\end{equation}
\begin{equation}
z_{2}=\frac{k_{2}}{2}+i(\omega _{2l}-\omega _{02}).
\end{equation}
When the external fields vanishes, the equations (\ref{R11M2}) and (\ref
{R22M2}) simplifies 
\begin{eqnarray}
\langle R_{11}(t)\rangle &=&\langle R_{11}(0)\rangle e^{-k_{1}t}+N(\omega
_{1})\left( 1-e^{-k_{1}t}\right) -  \nonumber \\
&&-2k_{1}N(\omega _{1})\int_{0}^{t}\langle R_{11}(t^{\prime })\rangle
e^{k_{1}\left( t^{\prime }-t\right) }dt^{\prime }-  \nonumber \\
&&-k_{1}N(\omega _{1})\int_{0}^{t}\langle R_{22}(t^{\prime })\rangle
e^{k_{1}\left( t^{\prime }-t\right) }dt^{\prime },
\end{eqnarray}
\begin{eqnarray}
\langle R_{22}(t)\rangle &=&\langle R_{22}(0)\rangle e^{-k_{2}t}+N(\omega
_{2})\left( 1-e^{-k_{2}t}\right) -  \nonumber \\
&&-2k_{2}N(\omega _{2})\int_{0}^{t}\langle R_{22}(t^{\prime })\rangle
e^{k_{2}\left( t^{\prime }-t\right) }dt^{\prime }-  \nonumber \\
&&-k_{2}N(\omega _{2})\int_{0}^{t}\langle R_{11}(t^{\prime })\rangle
e^{k_{2}\left( t^{\prime }-t\right) }dt^{\prime }.
\end{eqnarray}
Integrating these equations, we obtain 
\begin{eqnarray}
\langle R_{11}(t)\rangle &=&\langle R_{11}(0)\rangle e^{-\overline{k}_{1}t}+%
\frac{k_{1}}{\overline{k}_{1}}N(\omega _{1})\left( 1-e^{-\overline{k}%
_{1}t}\right) -  \nonumber \\
&&-\frac{k_{1}}{\overline{k}_{1}-\overline{k}_{2}}N(\omega _{1})\langle
R_{22}(0)\rangle \left( e^{-\overline{k}_{2}t}-e^{-\overline{k}_{1}t}\right),
\end{eqnarray}
\begin{eqnarray}
\langle R_{22}(t)\rangle &=&\langle R_{22}(0)\rangle e^{-\overline{k}_{2}t}+%
\frac{k_{2}}{\overline{k}_{2}}N(\omega _{2})\left( 1-e^{-\overline{k}%
_{2}t}\right) -  \nonumber \\
&&-\frac{k_{2}}{\overline{k}_{2}-\overline{k}_{1}}N(\omega _{2})\langle
R_{11}(0)\rangle \left( e^{-\overline{k}_{1}t}-e^{-\overline{k}_{2}t}\right),
\end{eqnarray}
where 
\begin{equation}
\overline{k}_{1}=k_{1}\left( 1+2N(\omega _{1})\right) ,
\end{equation}
\begin{equation}
\overline{k}_{2}=k_{2}\left( 1+2N(\omega _{2})\right).
\end{equation}
If the initial state of the system is given by
\begin{equation}
\psi=|0\rangle a_0 + |1\rangle a_1 + |2\rangle a_2
\end{equation}
with $|a_0|^2+|a_1|^2+|a_2|^2=1,$ then

\begin{eqnarray}
\langle R_{11}(t)\rangle &=&|a_1|^2 e^{-\overline{k}_{1}t}+%
\frac{k_{1}}{\overline{k}_{1}}N(\omega _{1})\left( 1-e^{-\overline{k}%
_{1}t}\right) -  \nonumber \\
&&-\frac{k_{1}}{\overline{k}_{1}-\overline{k}_{2}}N(\omega_{1})|a_2|^2%
\left(e^{-\overline{k}_{2}t}-e^{-\overline{k}_{1}t}\right),\label{R11l}
\end{eqnarray}

\begin{eqnarray}
\langle R_{22}(t)\rangle &=&|a_2|^2 e^{-\overline{k}_{2}t}+%
\frac{k_{2}}{\overline{k}_{2}}N(\omega _{2})\left( 1-e^{-\overline{k}%
_{2}t}\right) -  \nonumber \\
&&-\frac{k_{2}}{\overline{k}_{2}-\overline{k}_{1}}N(\omega _{2})
|a_1|^2 \left(
  e^{-\overline{k}_{1}t}-e^{-\overline{k}_{2}t}\right). \label{R22l} 
\end{eqnarray}

If the temperature of the bath is zero, and the system is initially in
one of the excited states, say, the state $|1\rangle$, then
$k_1=\overline{k}_1$, $a_0=a_2=0$ and 

   \begin{equation}
      \langle R_{11}(t) \rangle=|a_1|^2e^{-k_1t}
   \end{equation}
which is the expected spontaneous decay of the excited state with
lifetime $1/k_1$. If the temperature is different of zero, the thermal
fluctuations enhance the coupling between the system an the field and
the decay rate is increased, as we note by the equation
(\ref{R11l}). To times $t\gg 1/\bar k_1$, the system
approaches the saturation regime and then

   \begin{equation}
   \langle R_{11}(t=\infty)\rangle=\frac{N(\omega_1)}{1+2N(\omega_1)}.    
   \end{equation}
Assuming that the radiation field is in thermal equilibrium with the
cavity, we have 
\begin{equation}
  \label{eq:2}
  N(\omega_1)=\frac{1}{e^{h\omega_1/k_BT} -1} 
\end{equation}
and 
\begin{equation}
  \label{eq:4}
    N(\omega_2)=\frac{1}{e^{h\omega_1/k_BT} -1}.
\end{equation}
and the equations (\ref{R11l}) and (\ref{R22l}) becomes
\begin{eqnarray}
  \label{eq:3}
  \langle R_{11}(t=\infty)\rangle=\frac{1}{e^{h\omega_1/k_BT} +1}\\
  \langle R_{22}(t=\infty)\rangle=\frac{1}{e^{h\omega_2/k_BT} +1}
  \end{eqnarray}
which is the usual Fermi-Dirac distribution. 

\par With  the procedure discussed in detail in Appendix, we obtain the
solution of integral equations (\ref{R11M2}) and (\ref{R22M2}):
\begin{eqnarray}
\langle R_{11}(t)\rangle &=&\langle R_{11}(\infty )\rangle +\langle
R_{11}(0)\rangle \sum_{\substack{ i,j,k=0  \\ i\neq j\neq k 
}}^{3}\frac{\left|
\left( f_{i}+k_{1}/2\right) ^{2}+\Delta \omega _{1}^{2}\right| }{\left(
f_{i}-f_{j}\right) \left( f_{i}-f_{k}\right) }e^{f_{i}t}+  \nonumber \\
&&+k_1N(\omega _{1})\sum_{\substack{ i,j,k=0  \\ i\neq j\neq k 
}}^{3}\frac{\left|
\left( f_{i}+k_{1}/2\right) ^{2}+\Delta \omega _{1}^{2}\right| }{f_{i}\left(
f_{i}-f_{j}\right) \left( f_{i}-f_{k}\right) }e^{f_{i}t}+  \nonumber \\
&&+\frac{i\Omega _{1}\langle R_{10}(0)\rangle e^{i\phi }}{2}\sum_{\substack{ i,j,k=0 \\ i\neq j\neq k  }} ^{3}\frac{\left( f_{i}+z_{1}^{*}\right) }{\left(
f_{i}-f_{j}\right) \left( f_{i}-f_{k}\right) }e^{f_{i}t}-  \nonumber \\
&&-\frac{i\Omega _{1}\langle R_{01}(0)\rangle e^{-i\phi }}{2}\sum_{\substack{ i,j,k=0 \\ i\neq j\neq k  }} ^{3}\frac{\left( f_{i}+z_{1}\right) }{\left(
f_{i}-f_{j}\right) \left( f_{i}-f_{k}\right) }e^{f_{i}t}+  \nonumber \\
&&+\frac{\Omega _{1}^{2}}{2}\sum_{\substack{ i,j,k=0  \\ i\neq j\neq k 
}}^{3}%
\frac{\left( f_{i}+k_{1}/2\right) }{f_{i}\left( f_{i}-f_{j}\right) \left(
f_{i}-f_{k}\right) }e^{f_{i}t}-  \nonumber \\
&&-k_{1}N(\omega _{1})\langle R_{22}(0)\rangle \sum_{\substack{ i,j,k=0  \\ i\neq
j\neq k  }} ^{3}\frac{\left| \left( f_{i}+k_{1}/2\right) ^{2}+\Delta
\omega _{1}^{2}\right| \left| \left( f_{i}+k_{2}/2\right) ^{2}+\Delta \omega
_{2}^{2}\right| }{\left( f_{i}-f_{j}\right) \left( f_{i}-f_{k}\right)
h(f_{i})}e^{f_{i}t}-  \nonumber \\
&&-k_{1}N(\omega _{1})\langle R_{22}(0)\rangle \sum_{\substack{ i,j,k=0  \\ i\neq
j\neq k  }} ^{3}\frac{\left| \left( h_{i}+k_{1}/2\right) ^{2}+\Delta
\omega _{1}^{2}\right| \left| \left( h_{i}+k_{2}/2\right) ^{2}+\Delta \omega
_{2}^{2}\right| }{f(h_{i})\left( h_{i}-h_{j}\right) \left(
h_{i}-h_{k}\right) }e^{h_{i}t}-  \nonumber \\
&&-\Omega _{1}^{2}\langle R_{22}(0)\rangle \sum_{\substack{ i,j,k=0  \\ i\neq j\neq k 
}} ^{3}\frac{\left| \left( f_{i}+k_{2}/2\right) ^{2}+\Delta \omega
_{2}^{2}\right| \left( f_{i}+k_{1}/2\right) }{\left( f_{i}-f_{j}\right)
\left( f_{i}-f_{k}\right) h(f_{i})}e^{f_{i}t}-  \nonumber \\
&&-\Omega _{1}^{2}\langle R_{22}(0)\rangle \sum_{\substack{ i,j,k=0  \\ i\neq j\neq k 
}} ^{3}\frac{\left| \left( h_{i}+k_{2}/2\right) ^{2}+\Delta \omega
_{2}^{2}\right| \left( h_{i}+k_{1}/2\right) }{f(h_{i})\left(
f_{i}-f_{j}\right) \left( f_{i}-f_{k}\right) }e^{h_{i}t}.  \label{R11T}
\end{eqnarray}
and
\begin{eqnarray}
\langle R_{22}(t)\rangle &=&\langle R_{22}(\infty )\rangle +\langle
R_{22}(0)\rangle \sum_{\substack{ i,j,k=0  \\ i\neq j\neq k 
}}^{3}\frac{\left| \left(
h_{i}+k_{2}/2\right)^{2}+\Delta \omega _{2}^{2}\right| }{\left(
h_{i}-h_{j}\right) \left( h_{i}-h_{k}\right) }e^{h_{i}t}+  \nonumber \\
&&+k_{2}N(\omega _{2})\sum_{\substack{ i,j,k=0  \\ i\neq j\neq k 
}}^{3}\frac{%
\left| \left( h_{i}+k_{2}/2\right) ^{2}+\Delta \omega _{2}^{2}\right| }{%
h_{i}\left( h_{i}-h_{j}\right) \left( h_{i}-h_{k}\right) }e^{h_{i}t}+ 
\nonumber \\
&&+\frac{i\Omega _{2}\langle R_{20}(0)\rangle e^{i\phi }}{2}\sum_{\substack{ i,j,k=0 
\\ i\neq j\neq k  }} ^{3}\frac{\left( h_{i}+z_{2}^{*}\right) }{\left(
h_{i}-h_{j}\right) \left( h_{i}-h_{k}\right) }e^{h_{i}t}-  \nonumber \\
&&-\frac{i\Omega _{2}\langle R_{02}(0)\rangle e^{-i\phi }}{2}\sum_{\substack{ i,j,k=0 
\\ i\neq j\neq k  }} ^{3}\frac{\left( h_{i}+z_{2}\right) }{\left(
h_{i}-h_{j}\right) \left( h_{i}-h_{k}\right) }e^{h_{i}t}+  \nonumber \\
&&+\frac{\Omega _{2}^{2}}{2}\sum_{\substack{ i,j,k=0  \\ i\neq j\neq k 
}}^{3}%
\frac{\left( h_{i}+k_{2}/2\right) }{h_{i}\left( h_{i}-h_{j}\right) \left(
h_{i}-h_{k}\right) }e^{h_{i}t}-  \nonumber \\
&&-k_{2}N(\omega _{2})\langle R_{11}(0)\rangle \sum_{\substack{ i,j,k=0  \\ i\neq
j\neq k  }} ^{3}\frac{\left| \left( h_{i}+k_{2}/2\right) ^{2}+\Delta
\omega _{2}^{2}\right| \left| \left( h_{i}+k_{1}/2\right) ^{2}+\Delta \omega
_{1}^{2}\right| }{\left( h_{i}-h_{j}\right) \left( h_{i}-h_{k}\right)
f(h_{i})}e^{h_{i}t}-  \nonumber \\
&&-k_{2}N(\omega _{2})\langle R_{11}(0)\rangle \sum_{\substack{ i,j,k=0  \\ i\neq
j\neq k  }} ^{3}\frac{\left| \left( f_{i}+k_{2}/2\right) ^{2}+\Delta
\omega _{2}^{2}\right| \left| \left( f_{i}+k_{1}/2\right) ^{2}+\Delta \omega
_{1}^{2}\right| }{h(f_{i})\left( f_{i}-f_{j}\right) \left(
f_{i}-f_{k}\right) }e^{f_{i}t}-  \nonumber \\
&&-\Omega _{2}^{2}\langle R_{11}(0)\rangle \sum_{\substack{ i,j,k=0  \\ i\neq j\neq k 
}} ^{3}\frac{\left| \left( h_{i}+k_{1}/2\right) ^{2}+\Delta \omega
_{1}^{2}\right| \left( h_{i}+k_{2}/2\right) }{\left( h_{i}-h_{j}\right)
\left( h_{i}-h_{k}\right) f(h_{i})}e^{h_{i}t}-  \nonumber \\
&&-\Omega _{2}^{2}\langle R_{11}(0)\rangle \sum_{\substack{i,j,k=0 \\ i\neq j\neq k}}^{3}%
\frac{\left| \left( f_{i}+k_{1}/2\right) ^{2}+\Delta \omega _{1}^{2}\right|
\left( f_{i}+k_{2}/2\right) }{h(f_{i})\left( h_{i}-h_{j}\right) \left(
h_{i}-h_{k}\right) }e^{h_{i}t}.\label{R22T} 
\end{eqnarray}

In the above equations, we have used $\Delta \omega _{1}=\omega
_{1l}-\omega _{1}$, $\Delta \omega _{2}=\omega _{2l}-\omega _{2}$ and
\begin{equation}
\langle R_{11}(\infty )\rangle =\frac{\left( k_{1}^{2}/4+\Delta \omega
_{1}^{2}\right) N(\omega _{1})+\Omega _{1}^{2}/4}{\left( k_{1}^{2}/4+\Delta
\omega _{1}^{2}\right) \left( 1+2N(\omega _{1})\right) +\Omega _{1}^{2}/2},
\end{equation}
\begin{equation}
\langle R_{22}(\infty )\rangle =\frac{\left( k_{2}^{2}/4+\Delta \omega
_{2}^{2}\right) N(\omega _{2})+\Omega _{2}^{2}/4}{\left( k_{2}^{2}/4+\Delta
\omega _{2}^{2}\right) \left( 1+2N(\omega _{2})\right) +\Omega _{2}^{2}/2},
\end{equation}
\section{Discussion of the results}

\begin{figure}[p]
\psfrag{N(1)= 0}{$N(\omega_1)=0$}
\psfrag{N(1)= 0.1}{$N(\omega_1)=0.1$}
\psfrag{N(1)= 0.2}{$N(\omega_1)=0.2$}
\psfrag{N(1)= 0.5}{$N(\omega_1)=0.5$}
\psfrag{N(1)= 0.7}{$N(\omega_1)=0.7$}
\psfrag{N(1)= 1}{$N(\omega_1)=1$}
\psfrag{N(1)= 1.5}{$N(\omega_1)=1.5$}
\psfrag{R11}{$\langle R_{11}(\infty)\rangle$}
\psfrag{Delta1}{$\Delta\omega_1$}
\includegraphics[width=18cm]{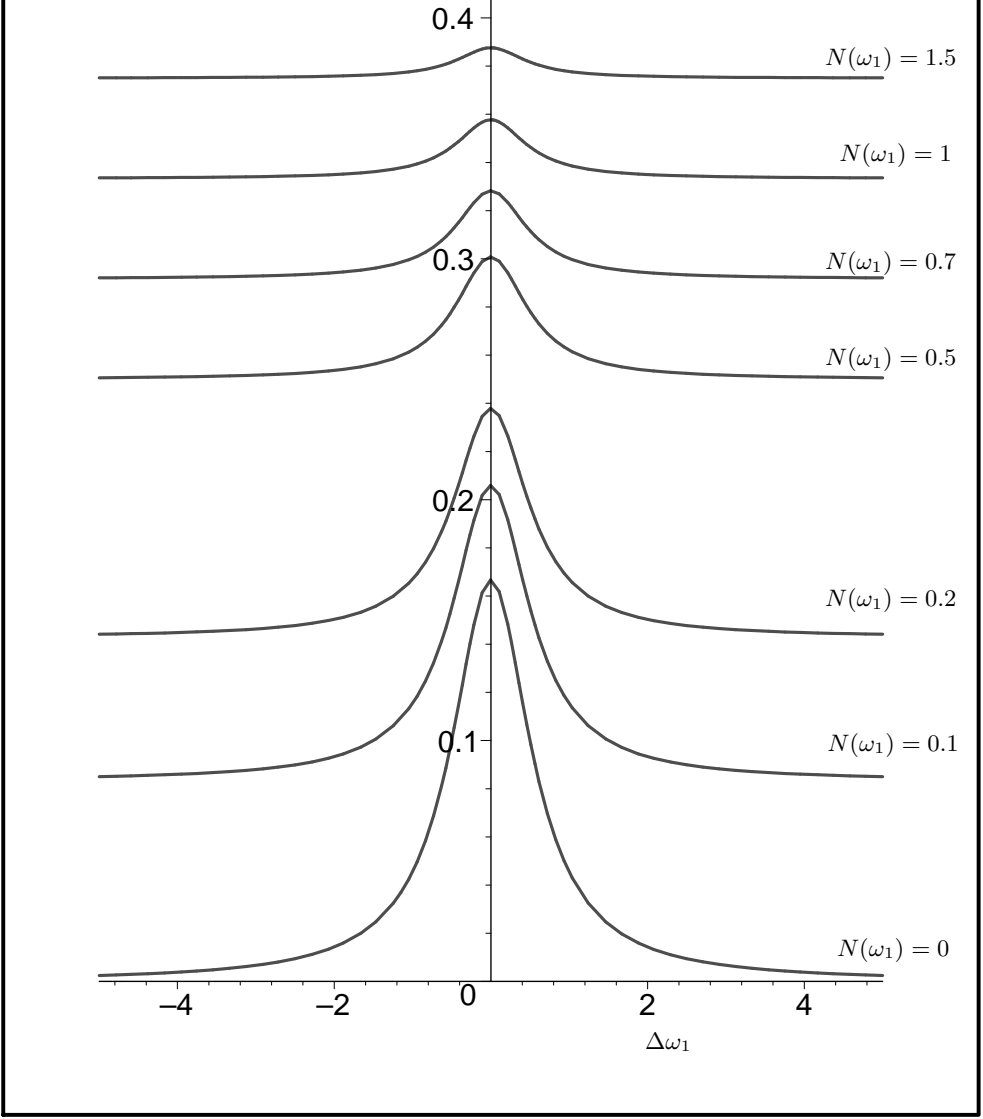}
\caption{$\langle R_{11}(\infty )\rangle $ dependence with the occupation
number $N(\omega_{1})$, the detuning $\Delta \omega_1$ and parameters
$\Omega_1=0.5$ and $k_1=1$.} \label{fig1}
\end{figure}

In  the equations (\ref{R11T})  and (\ref{R22T})  the only  terms that
remains when $t\rightarrow \infty$ are $\langle R_{11}(\infty)\rangle$
and  $\langle R_{22}(\infty)\rangle $.  The Fig.~\ref{fig1}  shows the
behavior of $\langle R_{11}(\infty)\rangle $ as a function of the 
detuning   $\Delta   \omega_{1}$   and   of  the   occupation   number
$N\left(w_{1}\right)$.  The  system   rapidly  tends  to  the  maximum
intensity    of     scattered    light    for    occupation    numbers
$N\left(w_{1}\right)  \neq 0$,  and  the peak  of  resonance with  the
driving  field   becomes  less  effective  to   increasing  number  of
occupation. 

\begin{figure}[p]
\psfrag{R11}{$\langle R_{11}(t)\rangle $}
\psfrag{R22}{$\langle R_{22}(t)\rangle $}
\psfrag{bt}{$\beta t$}
\includegraphics[width=18cm]{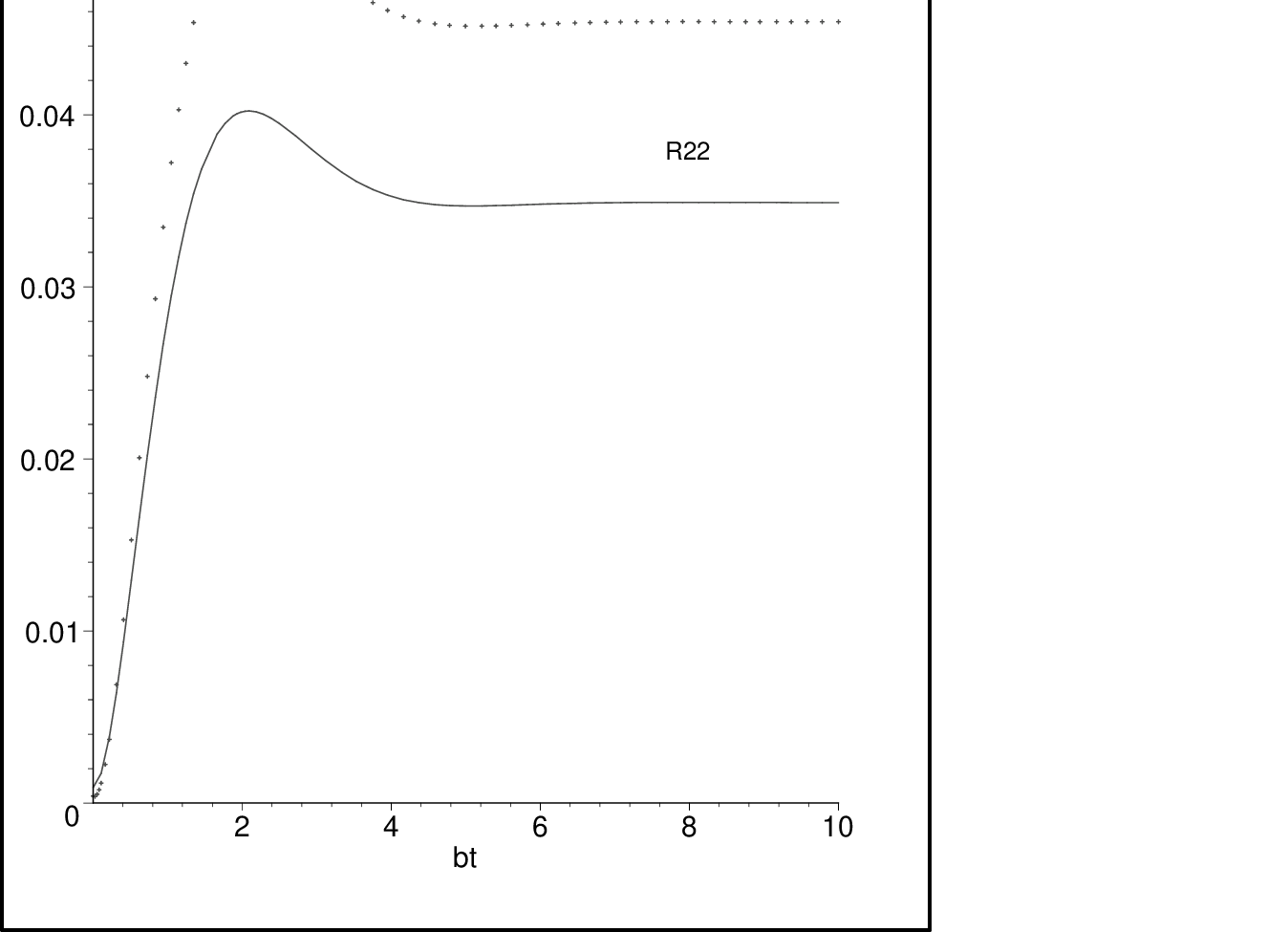}
\caption{Time development of the fluorescent light intensities $\langle
R_{11}(t)\rangle $ (points) and $\langle R_{22}(t)\rangle $ (solid line)
for Bose occupation numbers $N(w_{1})=N(w_{2})=0$ and $\lambda
_{1}=0.4$, $\lambda_{2}=0.3,$ $\theta _{1}=\theta _{2}=1$.}\label{fig2}
\end{figure}

\begin{figure}[p]
\psfrag{R11}{$\langle R_{11}(t)\rangle $}
\psfrag{R22}{$\langle R_{22}(t)\rangle $}
\psfrag{t}{$\beta t$}
\includegraphics[width=18cm]{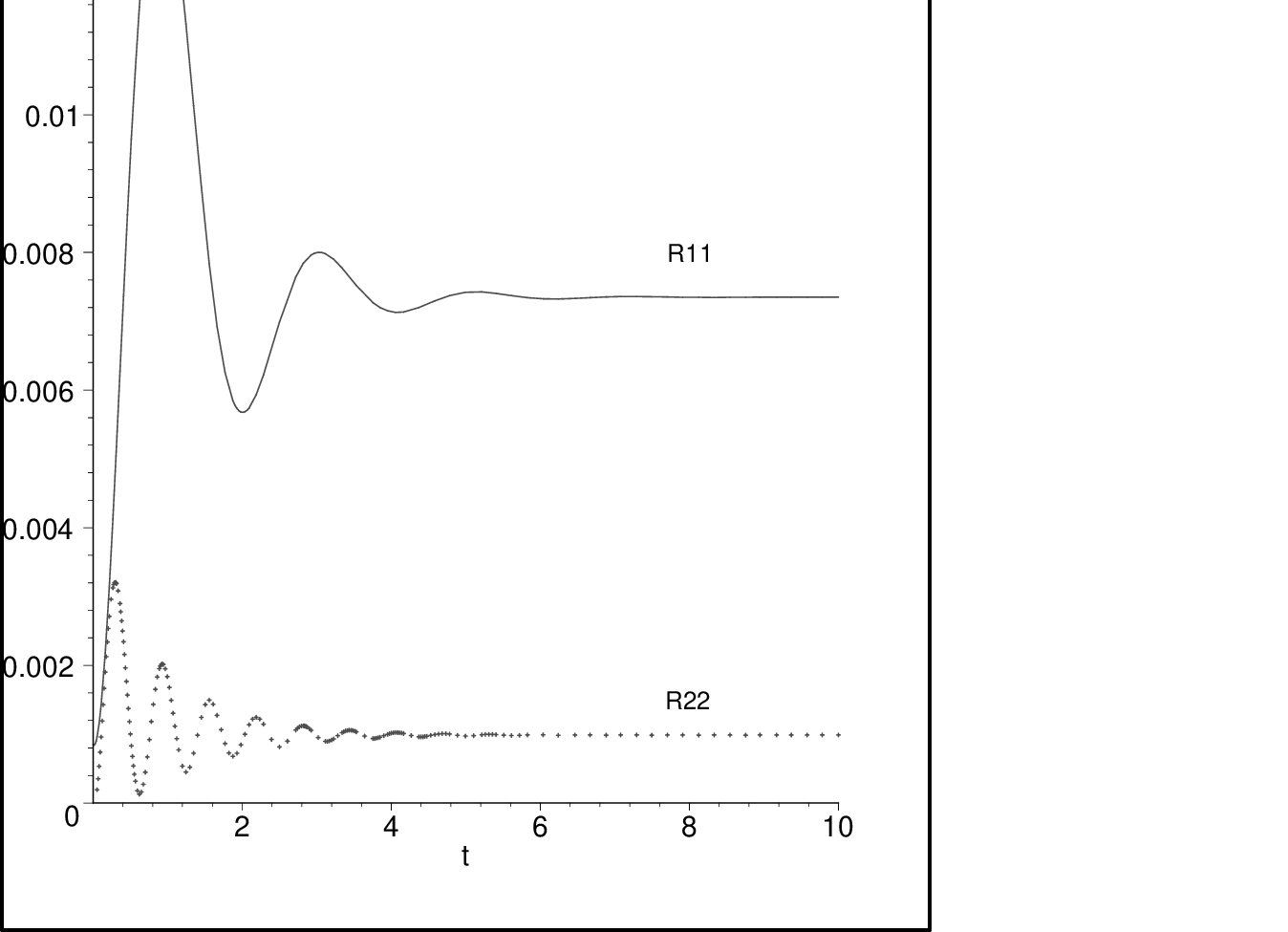}
\caption{Time development of the fluorescent light intensities $\langle
R_{11}(t)\rangle $ (points) and $\langle R_{22}(t)\rangle $ (solid line)
for Bose occupation numbers $N(w_{1})=N(w_{2})=0$ and $\lambda
_{1}=0.4$,    $\lambda_{2}=0.3$,   $\theta   _{1}=10$    and   $\theta
_{2}=3$.}\label{fig3} 
\end{figure}

\begin{figure}[p]
\psfrag{R11}{$\langle R_{11}(t)\rangle $}
\psfrag{R22}{$\langle R_{22}(t)\rangle $}
\psfrag{bt}{$\beta t$}
\includegraphics[width=18cm]{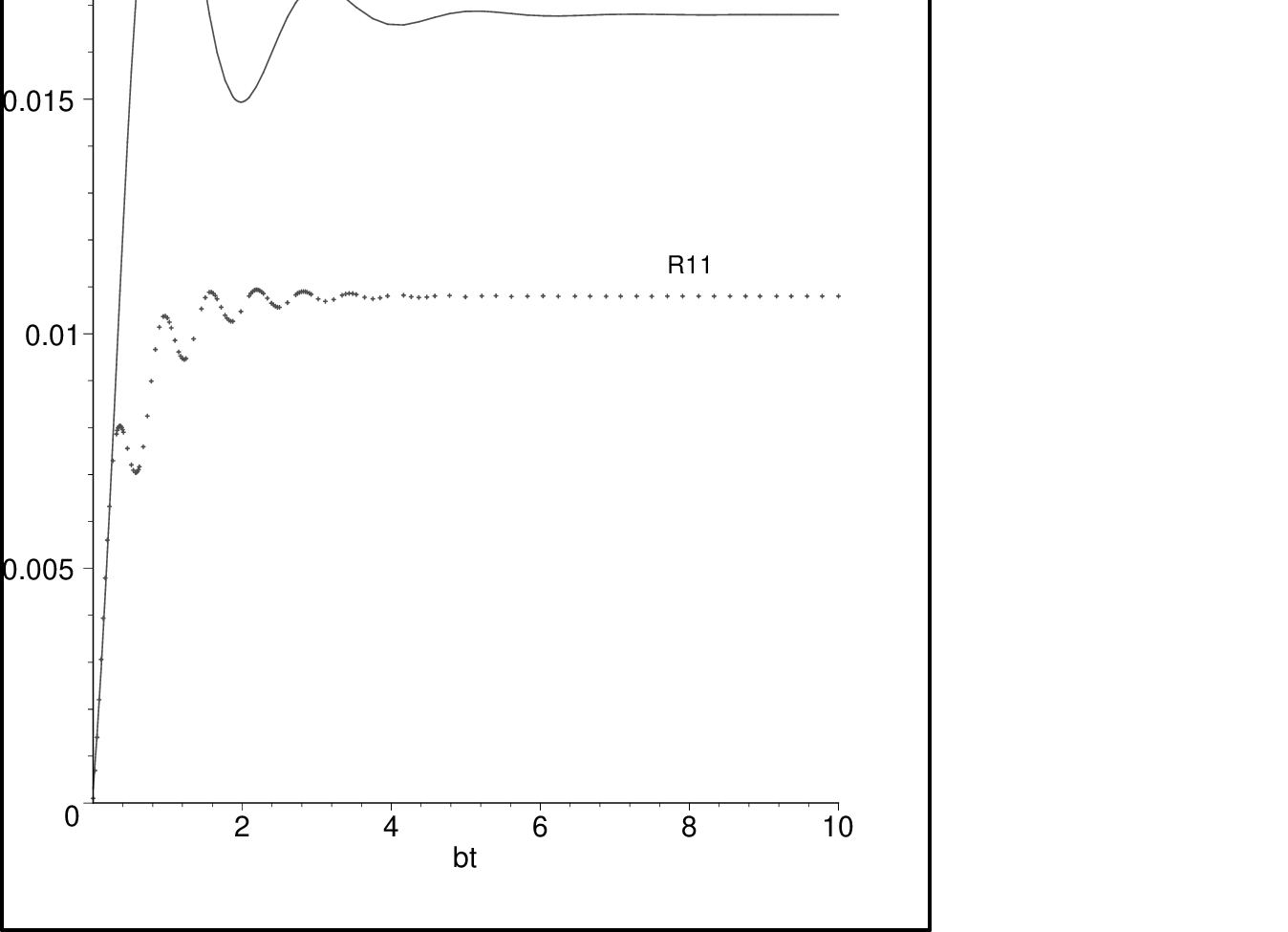}
\caption{Time development of the fluorescent light intensities $\langle
R_{11}(t)\rangle $ (points) and $\langle R_{22}(t)\rangle $ (solid line)
for Bose occupation numbers $N(w_{1})=N(w_{2})=0.01$ and $\lambda
_{1}=0.4$, $\lambda_{2}=0.3$, $\theta _{1}=10$ and $\theta _{2}=3$.} \label{fig4}
\end{figure}

\begin{figure}[p]
\psfrag{R11}{$\langle R_{11}(t)\rangle $}
\psfrag{R22}{$\langle R_{22}(t)\rangle $}
\psfrag{bt}{$\beta t$}
\includegraphics[width=18cm]{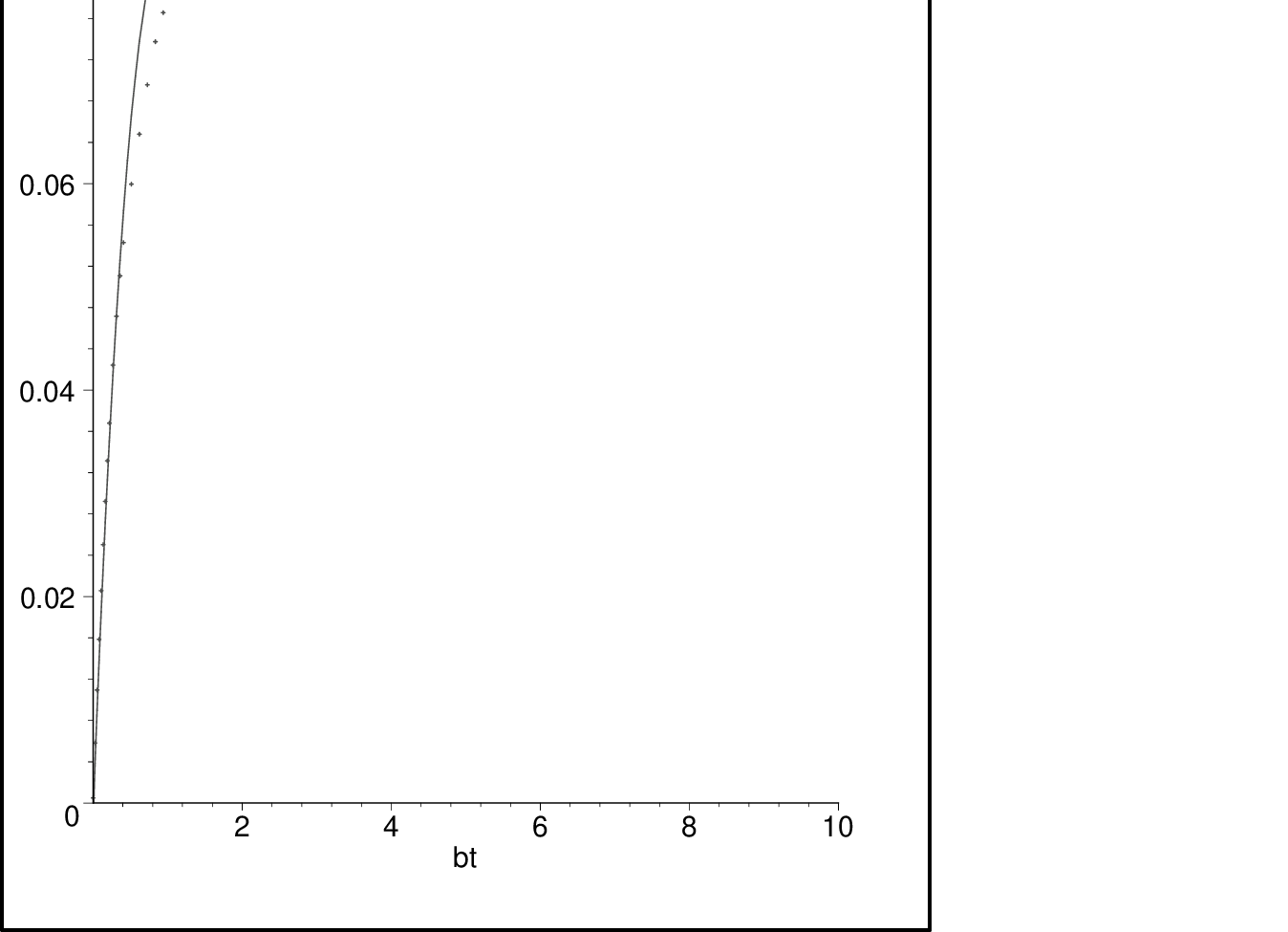}
\caption{Time development of the fluorescent light intensities $\langle
R_{11}(t)\rangle $ (points) and $\langle R_{22}(t)\rangle $ (solid line)
for Bose occupation numbers $N(w_{1})=N(w_{2})=0.1$ and $\lambda
_{1}=0.4$, $\lambda_{2}=0.3$, $\theta _{1}=10$ and $\theta _{2}=3$}\label{fig5}
\end{figure}

\begin{figure}[p]
\psfrag{N1=0}{$N(\omega_1)=0$}
\psfrag{N1=0.01}{$N(\omega_1)=0.01$}
\psfrag{N1=0.1}{$N(\omega_1)=0.1$}
\psfrag{bt}{$\beta t$}
\includegraphics[width=18cm]{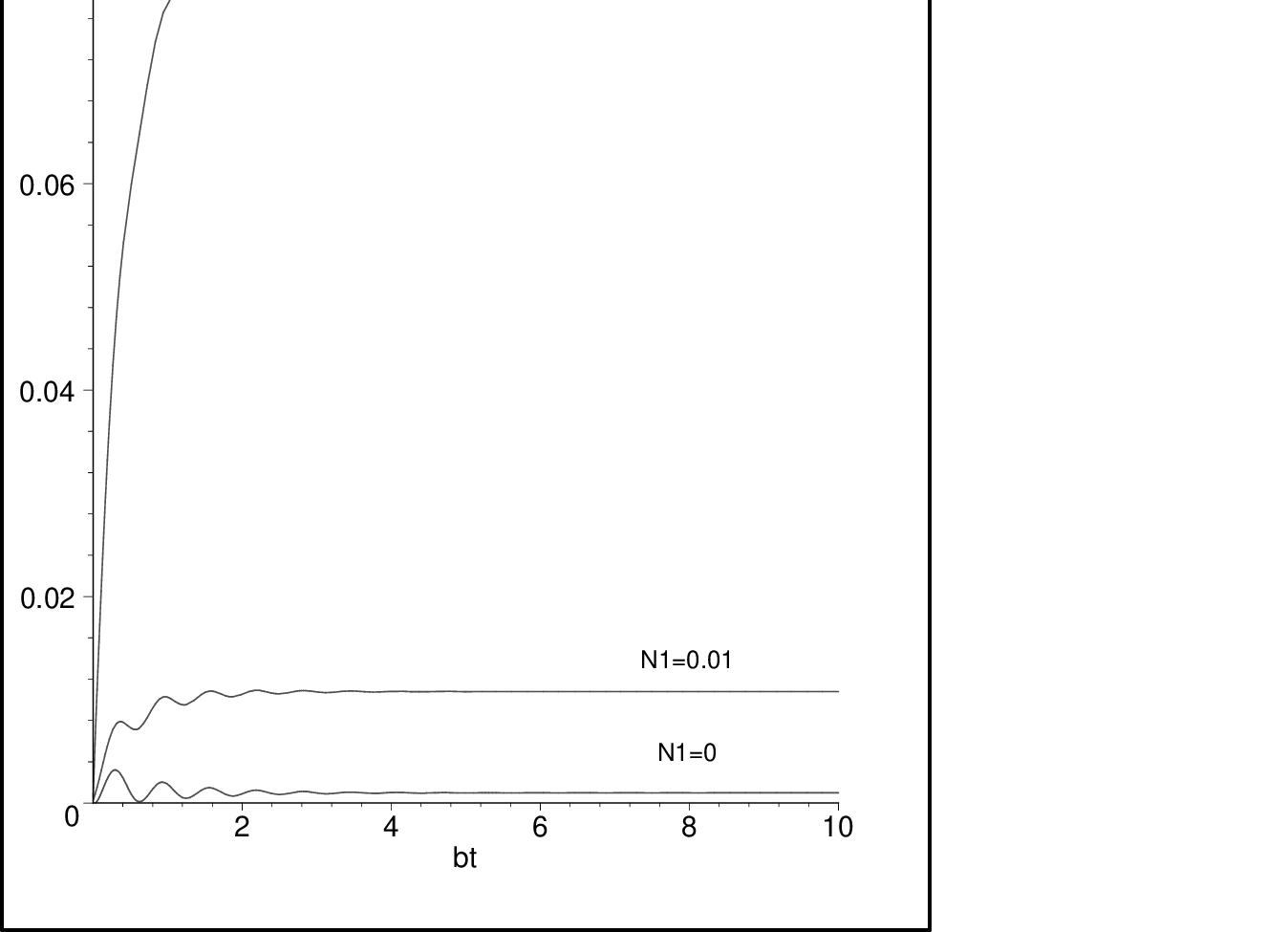}
\caption{Time development  of the fluorescent light  intensity $\langle
R_{11}(t)\rangle $ for $\lambda_1=0.4$, $\theta_1=10$, and increasing occupation numbers $N(\omega_1)$}.\label{fig6}
\end{figure}

\begin{figure}[p]
\psfrag{N1=0}{$N(\omega_1)=0$}
\psfrag{N1=0.01}{$N(\omega_1)=0.01$}
\psfrag{N1=0.1}{$N(\omega_1)=0.1$}
\psfrag{bt}{$\beta t$}
\includegraphics[width=18cm]{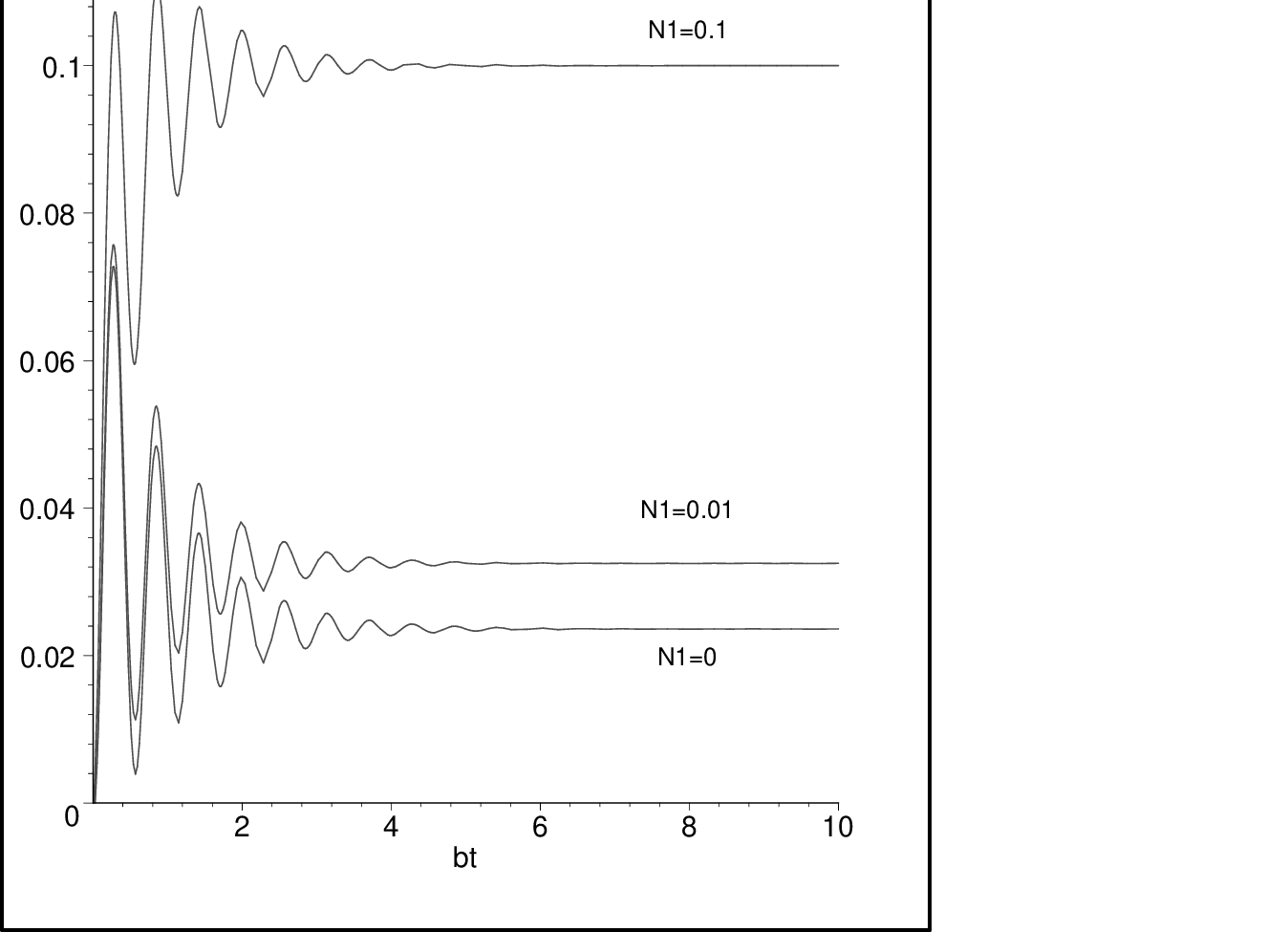}
\caption{Time development  of the fluorescent light  intensity $\langle
R_{11}(t)\rangle $ for $\lambda_1=10$, $\theta_1=10$, and increasing occupation numbers $N(\omega_1)$}\label{fig7}

\end{figure}

We see by an exam of the equations (\ref{R11T}) and (\ref{R22T}) that
one  excited  level  is  affected  by  the  other  only  in  transient
terms   that    depend   of    the   initial   values    of   $\langle
R_{11}(t=0)\rangle$, $\langle R_{22}(t=0)\rangle  $.  Thus, within the
second order approximation of the coupling constants we have used, the
upper levels are almost independents, and the 
three-level system  behaves as  two two-level systems.  Some numerical
values given by equation  (\ref{R11T}) for certain combinations of the
parameters $\lambda $ and $\theta $ for the initial 
state   $\langle   R_{11}(0)\rangle=\langle  R_{22}(0)\rangle=0$   and
$\langle R_{01}(0)\rangle =\langle  R_{10}(0)\rangle =0$, are given by
Figs.~\ref{fig2}~-~\ref{fig7}. With these initial values the equation
(\ref{R11T}) becomes
\begin{eqnarray}
\langle R_{11}(t)\rangle &=&\langle R_{11}(\infty )\rangle +k_1N(\omega _{1})\sum_{\substack{ i,j,k=0  \\ i\neq j\neq k 
}}^{3}\frac{\left|
\left( f_{i}+k_{1}/2\right) ^{2}+\Delta \omega _{1}^{2}\right| }{f_{i}\left(
f_{i}-f_{j}\right) \left( f_{i}-f_{k}\right) }e^{f_{i}t}+  \nonumber \\
&&+\frac{\Omega _{1}^{2}}{2}\sum_{\substack{ i,j,k=0  \\ i\neq j\neq k 
}}^{3}%
\frac{\left( f_{i}+k_{1}/2\right) }{f_{i}\left( f_{i}-f_{j}\right) \left(
f_{i}-f_{k}\right) }e^{f_{i}t} \label{eq:5}
\end{eqnarray}

The Fig.~\ref{fig2} shows the effect of
the   $\lambda_{1,2}$  parameters,  that   are  proportional   to  the
intensities  of  the  driving   fields,  over  the  fluorescent  light
intensities $\langle R_{11}(t)\rangle $ and $\langle R_{22}(t)\rangle$
for     Bose    occupation    numbers     $N(w_{1})=N(w_{2})=0$    and
$\theta_1=\theta_2=1$. 

In  the Fig.~\ref{fig3} the  light intensities
$\langle R_{11}(t)\rangle $  and $\langle R_{22}(t)\rangle$ are showed
to the  same parameters of  the Fig.~\ref{fig2}, but now  with $\theta
_{1}=10$ and  $\theta_{2}=3$.  It  can be seen  that the  evolution in
time of the fluorescent light intensity for $N\left( w_{1}\right) =0$
is increasing oscillatory as the detuning increases.  This feature is
attenuated for $N\left(\omega_1\right)\neq 0$, due to saturation caused
by the combined effect of the thermal bath and the driving fields.

The Fig.~\ref{fig4}  and Fig.~\ref{fig5} shows that a  increase in the
Bose occupation  numbers $N(w_{1})$  and $N(w_{2})$ increases  too the
light  intensities, but the  amplitude of  the oscillations  showed in
Fig.~\ref{fig3}  becomes  relatively   smaller.   The  system  rapidly
attains the thermal equilibrium with the bath, damping the oscillation
amplitude.  

The intensity of  the fluorescence light which for  weak driven fields
is determined  by the temperature  of the bath  (Fig.~\ref{fig6}), for
strong  fields depends  mainly of  the intensity  of the  driven field
(Fig.~\ref{fig7}).  

By the preceding examples, we see that the presence of the thermal
bath modifies strongly the time evolution of the fluorescent light
intensity. Thus the spectral density and  the intensity correlation  of
the  fluorescent light  must  too  be affected.  We  will present  these
calculations in a further paper.

\section{Appendix}
Equations (\ref{R11M2}) and (\ref{R22M2}) form a set of integral
equations that can be solved by Laplace transformation, 
   \begin{eqnarray}
\langle R_{11}(s)\rangle &=&\frac{\langle R_{11}(0)\rangle \left(
s+z_{1}\right) \left( s+z_{1}^{*}\right) }{f(s)}+\Gamma _{21}\frac{\left(
k_{1}-z_{1}\right) \left( s+z_{1}^{*}\right) }{f(s)}+  \nonumber \\
&&\frac{\Gamma _{21}^{*}\left( s+z_{1}\right) \left( k_{1}-z_{1}^{*}\right) 
}{f(s)}+\frac{k_{1}\Gamma _{11}\left( s+z_{1}\right) \left(
s+z_{1}^{*}\right) }{sf(s)}-  \nonumber \\
&&-\frac{\langle R_{22}(s)\rangle }{f(s)}\left[ k_{1}N(\omega _{1})\left(
s+z_{1}\right) \left( s+z_{1}^{*}\right) +\Omega _{1}^{2}\left(
s+k_{1}/2\right) \right].  \nonumber \\
&&  \label{R11L}
   \end{eqnarray}
   \begin{eqnarray}
\langle R_{22}(s)\rangle &=&\frac{\langle R_{22}(0)\rangle \left(
s+z_{2}\right) \left( s+z_{2}^{*}\right) }{h(s)}+\Gamma _{22}\frac{\left(
k_{2}-z_{2}\right) \left( s+z_{2}^{*}\right) }{h(s)}+  \nonumber \\
&&\frac{\Gamma _{22}^{*}\left( s+z_{2}\right) \left( k_{1}-z_{2}^{*}\right) 
}{h(s)}+\frac{k_{2}\Gamma _{12}\left( s+z_{2}\right) \left(
s+z_{2}^{*}\right) }{sh(s)}-  \nonumber \\
&&-\frac{\langle R_{11}(s)\rangle }{h(s)}\left[ k_{2}N(\omega _{2})\left(
s+z_{2}\right) \left( s+z_{2}^{*}\right) +\Omega _{2}^{2}\left(
s+k_{2}/2\right) \right] ,  \nonumber \\
&&  \label{R22L}
   \end{eqnarray}
In equations $\left( \text{\ref{R11L}}\right) $ and (\ref{R22L}) we have
used 
   \begin{equation}
f(s)=\left( s+k_{1}\right) \left( s+z_{1}\right) \left( s+z_{1}^{*}\right)
+2k_{1}N(\omega _{1})\left( s+z_{1}\right) \left( s+z_{1}^{*}\right) +\Omega
_{1}^{2}\left( s+k_{1}/2\right) ,
   \end{equation}
   \begin{equation}
h(s)=\left( s+k_{2}\right) \left( s+z_{2}\right) \left( s+z_{2}^{*}\right)
+2k_{2}N(\omega _{2})\left( s+z_{2}\right) \left( s+z_{2}^{*}\right) +\Omega
_{2}^{2}\left( s+k_{2}/2\right)
   \end{equation}
Substituting (\ref{R22L}) in (\ref{R11L}), and retaining the terms until
second order in the coupling constants, we have 
   \begin{eqnarray}
\langle R_{11}(s)\rangle &=&\frac{\langle R_{11}(0)\rangle \left(
s+z_{1}\right) \left( s+z_{1}^{*}\right) }{f(s)}+\frac{\Gamma
_{21}z_{1}^{*}\left( s+z_{1}^{*}\right) }{f(s)}+  \nonumber \\
&&\frac{\Gamma _{21}^{*}z_{1}\left( s+z_{1}\right) }{f(s)}+\frac{k_{1}\Gamma
_{11}\left( s+z_{1}\right) \left( s+z_{1}^{*}\right) }{sf(s)}-  \nonumber \\
&&-\frac{\langle R_{22}(0)\rangle \left( s+z_{2}\right) \left(
s+z_{2}^{*}\right) }{f(s)h(s)}\left[ k_{1}N(\omega _{1)}\left(
s+z_{1}\right) \left( s+z_{1}^{*}\right) +\Omega _{1}^{2}\left(
s+k_{1}/2\right) \right].  \nonumber
   \end{eqnarray}
Taking now the inverse Laplace transform of the above equation, 
   \begin{equation}
\langle R_{11}(t)\rangle =\frac{1}{2\pi i}\int_{-i\infty +\alpha }^{i\infty
+\alpha }\langle R_{11}(s)\rangle e^{st}ds,\qquad \;t>0
\end{equation}  
where $\alpha $ is chosen in such a way that all singularities of the
integrand fall in  the left of the line Re $s=\alpha  $ in the complex
plane. Using the residue theorem, we obtain the 
wanted integral over the line Re $s=\alpha $. Assuming that we may write 
   \begin{equation}
   f(s)=\ \left( s-f_{1}\right) \left( s-f_{2}\right) \left( s-f_{3}\right) ,
   \end{equation}
   \begin{equation}
   h(s)=\ \left( s-h_{1}\right) \left( s-h_{2}\right) \left( s-h_{3}\right) ,
   \end{equation}
where $f_{1},$ $f_{2},$ $f_{3}$ are the three roots of the cubic equation 
\[
s^{3}+\left[ 1+N(w_{1})\right] 2k_{1}s^{2}+\left| \left[ 5+8N\left(
w_{1}\right) \right] +\theta _{1}^{2}+\lambda _{1}\right| \beta _{1}^{2}s+ 
\]
   \begin{equation}
   +\left| 2\left[ 1+2N\left( w_{1}\right) \right] +2\left[ 1+2N\left(
   w_{1}\right) \right] \beta _{1}^{2}+\lambda _{1}\right| \beta _{1}^{3}=0,
   \end{equation}
and $h_{1},$ $h_{2},$ $h_{3}$ are the three roots of the cubic equation 
\[
s^{3}+\left[ 1+N(w_{2})\right] 2k_{2}s^{2}+\left| \left[ 5+8N\left(
w_{2}\right) \right] +\theta _{2}^{2}+\lambda _{2}\right| \beta _{2}^{2}s+ 
\]
   \begin{equation}
+\left| 2\left[ 1+2N\left( w_{2}\right) \right] +2\left[ 1+2N\left(
w_{2}\right) \right] \beta _{2}^{2}+\lambda _{2}\right| \beta _{2}^{3}=0,
   \end{equation}
where $\theta _{i}=2\Delta \omega _{i}/k_{i}$ , $\lambda _{i}=\Omega
_{i}^{2}/\beta _{i}^{2}$ , $\beta _{i}=k_{i}/2$ , with $i=1,2$. Then we have 
\begin{eqnarray}
\langle R_{11}(t)\rangle &=&\langle R_{11}(\infty )\rangle +\langle
R_{11}(0)\rangle \sum_{\substack{ i,j,k=0  \\ i\neq j\neq k 
}}^{3}\frac{\left|
\left( f_{i}+k_{1}/2\right) ^{2}+\Delta \omega _{1}^{2}\right| }{\left(
f_{i}-f_{j}\right) \left( f_{i}-f_{k}\right) }e^{f_{i}t}+  \nonumber \\
&&+kN(\omega _{1})\sum_{\substack{ i,j,k=0  \\ i\neq j\neq k 
}}^{3}\frac{\left|
\left( f_{i}+k_{1}/2\right) ^{2}+\Delta \omega _{1}^{2}\right| }{f_{i}\left(
f_{i}-f_{j}\right) \left( f_{i}-f_{k}\right) }e^{f_{i}t}+  \nonumber \\
&&+\frac{i\Omega _{1}\langle R_{10}(0)\rangle e^{i\phi }}{2}\sum_{\substack{ i,j,k=0 \\ i\neq j\neq k  }} ^{3}\frac{\left( f_{i}+z_{1}^{*}\right) }{\left(
f_{i}-f_{j}\right) \left( f_{i}-f_{k}\right) }e^{f_{i}t}-  \nonumber \\
&&-\frac{i\Omega _{1}\langle R_{01}(0)\rangle e^{-i\phi }}{2}\sum_{\substack{ i,j,k=0 \\ i\neq j\neq k  }} ^{3}\frac{\left( f_{i}+z_{1}\right) }{\left(
f_{i}-f_{j}\right) \left( f_{i}-f_{k}\right) }e^{f_{i}t}+  \nonumber \\
&&+\frac{\Omega _{1}^{2}}{2}\sum_{\substack{ i,j,k=0  \\ i\neq j\neq k 
}}^{3}%
\frac{\left( f_{i}+k_{1}/2\right) }{f_{i}\left( f_{i}-f_{j}\right) \left(
f_{i}-f_{k}\right) }e^{f_{i}t}-  \nonumber \\
&&-k_{1}N(\omega _{1})\langle R_{22}(0)\rangle \sum_{\substack{ i,j,k=0  \\ i\neq
j\neq k  }} ^{3}\frac{\left| \left( f_{i}+k_{1}/2\right) ^{2}+\Delta
\omega _{1}^{2}\right| \left| \left( f_{i}+k_{2}/2\right) ^{2}+\Delta \omega
_{2}^{2}\right| }{\left( f_{i}-f_{j}\right) \left( f_{i}-f_{k}\right)
h(f_{i})}e^{f_{i}t}-  \nonumber \\
&&-k_{1}N(\omega _{1})\langle R_{22}(0)\rangle \sum_{\substack{ i,j,k=0  \\ i\neq
j\neq k  }} ^{3}\frac{\left| \left( h_{i}+k_{1}/2\right) ^{2}+\Delta
\omega _{1}^{2}\right| \left| \left( h_{i}+k_{2}/2\right) ^{2}+\Delta \omega
_{2}^{2}\right| }{f(h_{i})\left( h_{i}-h_{j}\right) \left(
h_{i}-h_{k}\right) }e^{h_{i}t}-  \nonumber \\
&&-\Omega _{1}^{2}\langle R_{22}(0)\rangle \sum_{\substack{ i,j,k=0  \\ i\neq j\neq k 
}} ^{3}\frac{\left| \left( f_{i}+k_{2}/2\right) ^{2}+\Delta \omega
_{2}^{2}\right| \left( f_{i}+k_{1}/2\right) }{\left( f_{i}-f_{j}\right)
\left( f_{i}-f_{k}\right) h(f_{i})}e^{f_{i}t}-  \nonumber \\
&&-\Omega _{1}^{2}\langle R_{22}(0)\rangle \sum_{\substack{ i,j,k=0  \\ i\neq j\neq k 
}} ^{3}\frac{\left| \left( h_{i}+k_{2}/2\right) ^{2}+\Delta \omega
_{2}^{2}\right| \left( h_{i}+k_{1}/2\right) }{f(h_{i})\left(
f_{i}-f_{j}\right) \left( f_{i}-f_{k}\right) }e^{h_{i}t}. 
\end{eqnarray}

With a similar procedure we obtain a solution to $\langle R_{22}(t)\rangle $.


\newpage

\listoffigures

\end{document}